\documentclass[pra,aps,twocolumn,showpacs,10pt]{revtex4-1}
\usepackage{amsmath,amssymb,graphicx}
\usepackage{color}
\usepackage{subfigure}

\begin{document}

\title{Rotating azimuthons in dissipative Kerr media excited by
superpositions of Bessel beams}
\author{Carlos Ruiz-Jim\'enez$^1$}
\author{Herv\'e Leblond$^2$}
\author{Miguel A. Porras$^1$}
\author{Boris A. Malomed$^{3,4}$}
\affiliation{$^1$Grupo de Sistemas Complejos, ETSIME, Universidad
Polit\'ecnica de Madrid, Rios Rosas 21, 28003 Madrid, Spain}
\affiliation{$^2$Laboratoire de Photonique d'Angers, EA 4464, Universit\'e
d'Angers, 2 Bd Lavoisier, 49000 Angers, France}
\affiliation{$^3$Department
of Physical Electronics, School of Electrical Engineering, Faculty of
Engineering, and Center for Light-Matter interaction, Tel Aviv University,
Tel Aviv 69978, Israel}
\affiliation{$^4$Instituto de Alta
Investigaci\'{o}n, Universidad de Tarapac\'{a}, Casilla 7D, Arica, Chile}

\begin{abstract}
We report the existence of persistently rotating azimuthons in media with
self-focusing Kerr and absorption nonlinearities. The nonlinear loss is
balanced by power influx from the peripheral reservoir stored in a slowly
decaying tail of the field. The azimuthon modes are excited by a
superposition of two Bessel beams with opposite vorticities, $\pm s$, and
slightly different conicities. The excited mode exhibits vorticity in its
center opposite to that of the input Bessel-beam superposition, due to
spontaneous inversion of the topological charge in the course of the
azimuthon formation. Unlike azimuthons in loss-free media, number $N$ of
rotating intensity maxima and $s$ are not mutually independent, being
related by $N=2s$. The robustness of the rotating azimuthons is enhanced in
comparison to similar static dissipative patterns. They can be excited in
almost any transparent material, in the range of intensities for which the
nonlinear absorption, induced by multiphoton absorption, is relevant. Close
to the ionization threshold, the rotating azimuthons are similar to recently
observed helical filaments of light in air and CS$_{2}$.
\end{abstract}

\maketitle

\section{Introduction}

Solitons, solitary vortices \cite{REVIEW,BORIS}, necklace-shaped clusters
\cite{DESYATNIKOV1,KARTASHOV}, and azimuthons \cite{DESYATNIKOV2,MINO} are
increasingly sophisticated, self-trapped light modes in nonlinear optical
media, which have been predicted and experimentally realized in the course
of the last decades \cite{review}. In particular, soliton clusters and
azimuthons feature propagation-invariant or nearly-invariant intensity
patterns that rotate uniformly as they propagate in transparent media, and
require a stabilizing mechanism to arrest the collapse instability brought
by the self-focusing Kerr nonlinearity. Such a mechanism may be provided by
saturation of the Kerr nonlinearity \cite{DESYATNIKOV1,DESYATNIKOV2,ICFO}.
Many of these solitary structures have dissipative-soliton counterparts,
which are supported by the balance between gain and losses, which occurs in
laser cavities, in addition to the balance between the diffraction and
self-focusing \cite{AKHMEDIEV,NNR1,NNR2,Thawatchai}.

Relaxing the condition of strong localization, one can consider weakly
localized states similar to Bessel beams \cite%
{PORRASPRL,POLESANA,POLESANABESSEL}, whose total norm (integral power, in
terms of optics) diverges at $r\rightarrow \infty ,$ where $r$ is the radial
coordinate in the two-dimensional plane, perpendicular to the propagation
axis, $z$. Unlike dissipative solitons, stationarity of such states is
supported not through the balance of loss and gain, but rather due to the
compensation of nonlinear dissipation, induced by multiphoton absorption in
the optical material (which may generate weak plasma) and influx of power
stored, in an indefinitely large (diverging) amount, in the weakly decaying
tail of the quasi-Gaussian beam \cite%
{PORRASPRL,POLESANA,POLESANABESSEL,COUAIRON}.

Similarly to the above-mentioned conservative systems, nonlinear dissipative
Bessel beams with embedded vorticity have been predicted \cite%
{PORRASJOSAB,JUKNA}, and experimentally observed to induce tubular
filamentation \cite{XIU}. A remarkable fact is that the dissipative
nonlinear Bessel vortex beams can be stable in self-focusing media with the
pure-cubic (Kerr) nonlinearity due to the stabilizing action of the
nonlinear absorption \cite{PORRASPRA1}. More recently, launching arbitrary
superpositions of Bessel beams with the same cone angle but different
embedded vorticities has been shown to excite propagation-invariant
(stationary) dissipation patterns of rather arbitrary shapes, the so-called
``dissipatons" \cite{PORRASPRA2}.

In this work, which is motivated, in part, by recent experiments exhibiting
helical filamentation of light beams \cite{BARBIERI,LU}, we predict the
existence of what we call \textit{rotating dissipative azimuthon}s. These
modes propagate steadily, with a constant rotation velocity, in nonlinearly
absorbing Kerr media, being excited by superpositions of two Bessel vortex
beams with opposite vorticities and slightly different conicities. Such
coherent superpositions of Bessel beams can be readily generated in the
experiment. In the linear approximation, their propagation was theoretically
studied in Refs. \cite{VASILYEU,ROP,PORRASJOSAA}.

Unlike soliton clusters and azimuthons in conservative systems, number $N$
of rotating high-intensity peaks (``hot spots") and
vorticity $s$ at the center of the dissipative azimuthon are not independent
integers, but are related by $N=2s$, which can be explained analytically
(see below).
The rotating azimuthons exhibit a mixed linear-nonlinear behavior,
resembling in some aspects the linear propagation of superpositions of
Bessel beams with opposite vorticities and different cone angles, while in
other respects they are similar to solitons clusters. Namely, their
quasi-linear peripheral field imposes the same angular velocity of rotation
as that induced by the Bessel-beam superposition, see Eq. (\ref{Omega})
below. However, the vorticity at the center of the azimuthon acquires a
topological charge \emph{opposite} to that of the input superposition of the
Bessel wave functions, as a result of the topological-charge inversion in
the course of the formation of the azimuthon, which can be explained by a
trend to minimization of the Hamiltonian of the model's conservative part.
As a result, the dissipative azimuthon rotates in the direction opposite to
the azimuthal gradient of the phase associated with the vorticity at the
center, in the same way as soliton clusters do.

We also report a gyroscopic effect in the spontaneous formation of
dissipative azimuthons, \textit{viz}., that the rotation accelerates the
formation of the spinning steady-shape patterns, in comparison to similar
non-rotating steady ones (``dissipatons"), which arise when
cone angles of the two Bessel beams are equal in the input state. We also
observe enhanced stability of the rotating dissipative azimuthons, as
compared to dissipative Bessel vortex beams and static dissipatons. In the
case of instability, the rotating dissipatons feature richer dynamics,
including formation of persistently pulsating azimuthons.

Basic results, produced by systematic numerical simulations of the model,
are reported in Section II. The gyroscopic effect and a detailed analysis of
the stability of the rotating dissipative azimuthons are presented in
Section III and IV, respectively.
The paper is concluded by
Section V, while Appendix presents specific details of the numerical
algorithms used in the paper.

\section{Dissipative azimuthons excited by superpositions of Bessel beams
with opposite topological charges}

We consider the propagation of monochromatic light beams along the $z$ axis,
$E=A\exp (-i\omega t+ikz)$, with carrier frequency $\omega $ and propagation
constant $k=n\omega /c$, where $c$ is the speed of light in vacuum, and $n$
is the linear refractive index. The nonlinear Schr\"{o}dinger equation
(NLSE) that governs the paraxial propagation of field envelope $A$, is \cite%
{PORRASJOSAB}
\begin{equation}
\partial _{z}A=\frac{i}{2k}\nabla _{\perp }^{2}A+\frac{ikn_{2}}{n}|A|^{2}A-%
\frac{\beta ^{(M)}}{2}|A|^{2M-2}A\,,  \label{NLSE}
\end{equation}%
where $\nabla _{\perp }^{2}\equiv \partial _{r}^{2}+(1/r)\partial
_{r}+(1/r^{2})\partial _{\varphi }^{2}$ is the transverse Laplacian, which
is here written in polar coordinates $(r,\varphi )$ in the transverse plane
(all simulations were performed, in parallel, in both the Cartesian and polar coordinates, see
Appendix), $n_{2}>0$ is the nonlinear refractive index, and $\beta ^{(M)}>0$
is the multiphoton absorption coefficient of order $M$.

\begin{figure}[tbp]
\includegraphics*[width=4.0cm]{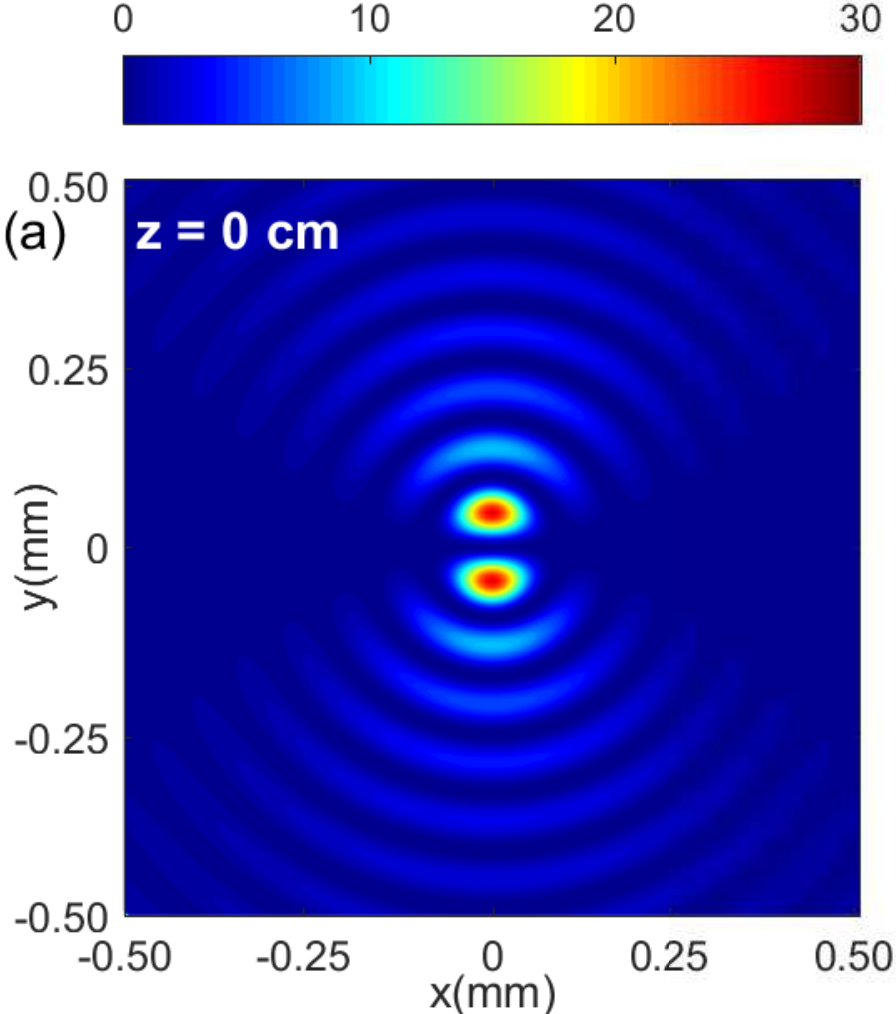} \includegraphics*[
width=4.0cm]{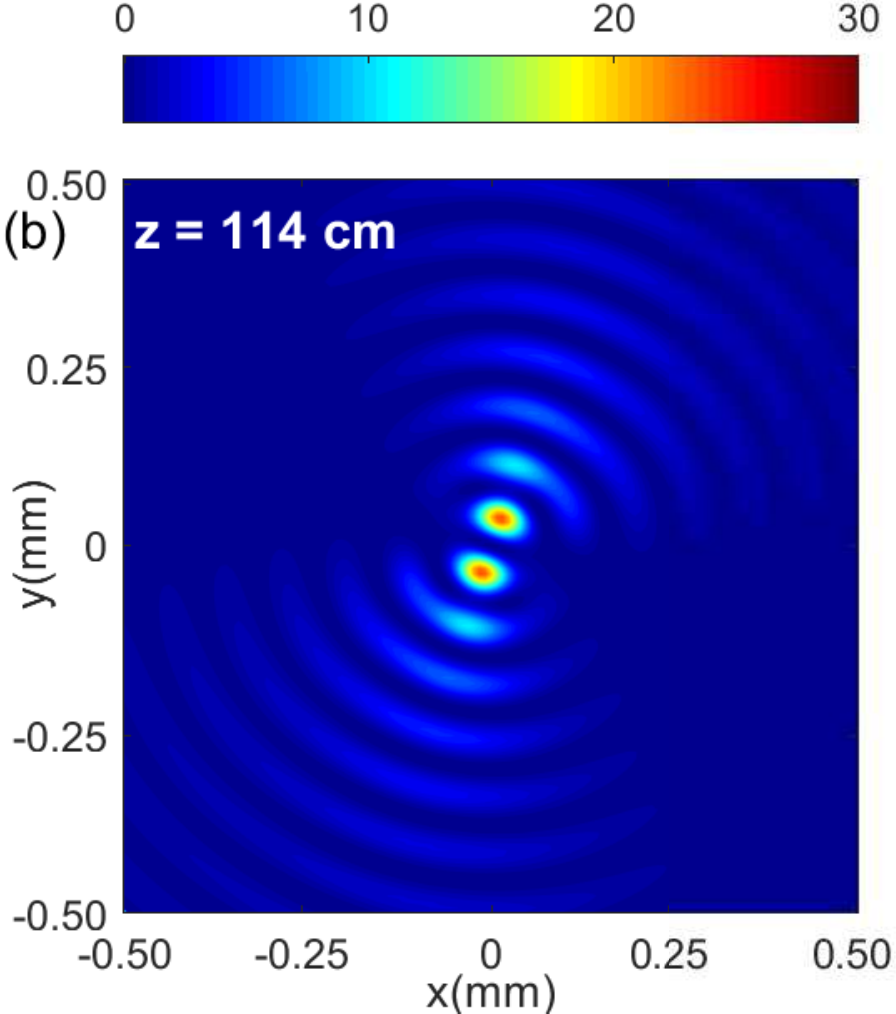} \includegraphics*[width=4.0cm]{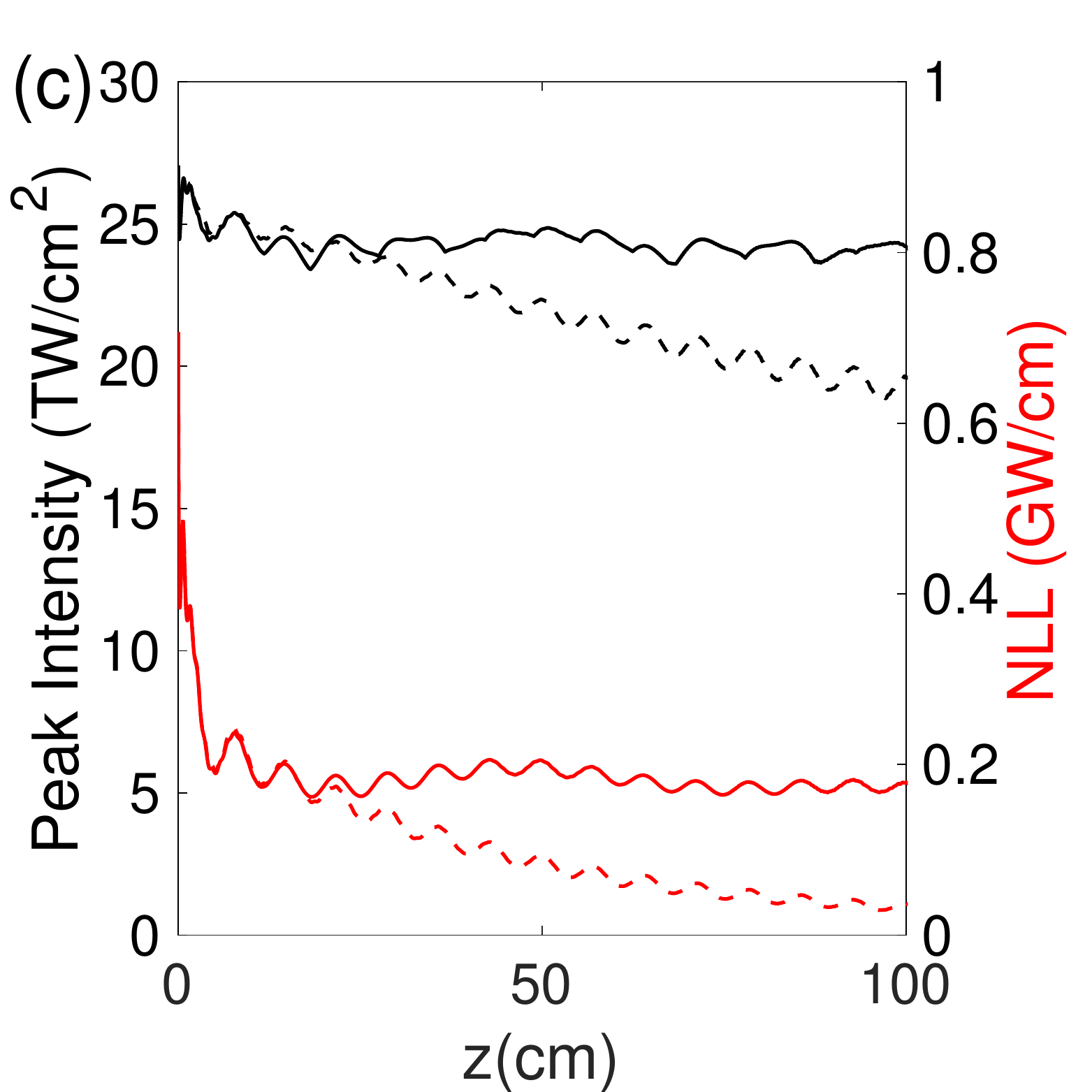} %
\includegraphics*[width=4.0cm]{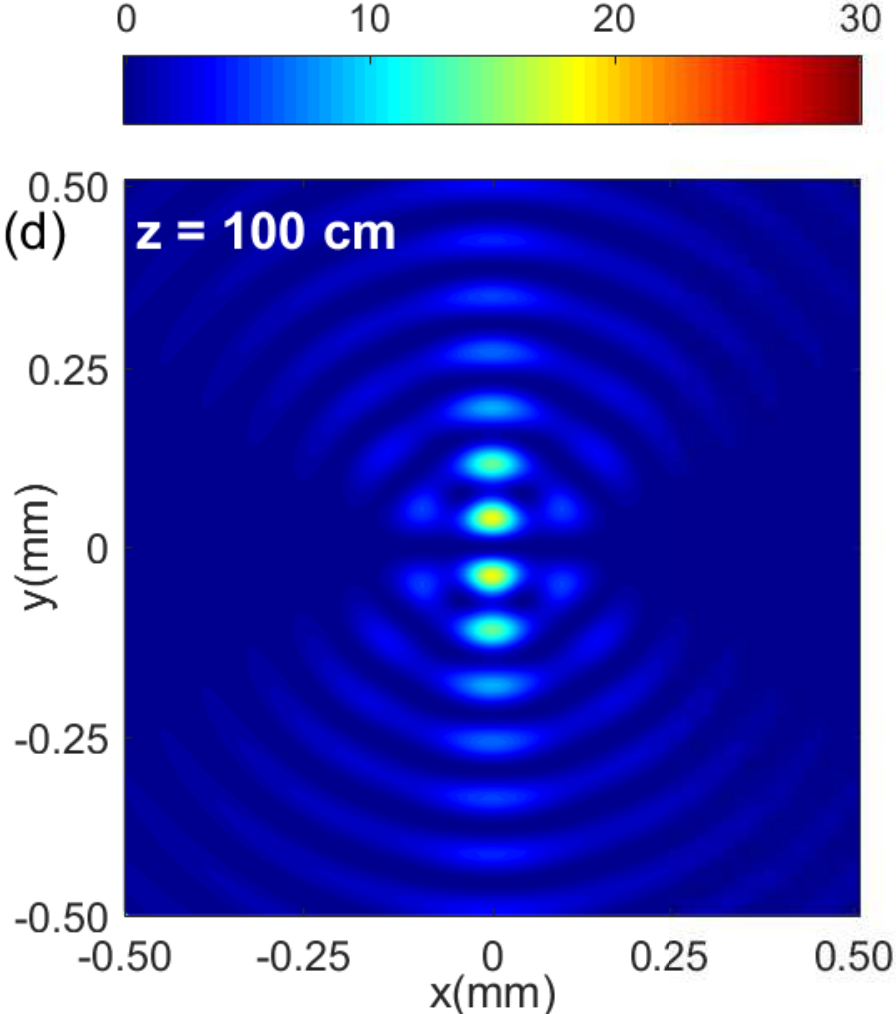}
\caption{Results of simulations of Eq. (\protect\ref{NLSE}) with parameters
corresponding to the high-intensity light propagation in air at the carrier
wavelength $800$ nm ($n\approx 1$, $n_{2}=3.2\times 10^{-19}$ cm$^{2}$/W, $%
M=8$, $\protect\beta ^{(8)}=1.8\times 10^{-94}$ cm$^{13}$/W$^{7}$
\protect\cite{COUAIRON}). (a) The input intensity profile, given by Eq.
\eqref{BESSEL-AZIMUTHON}, and measured in units of TW/cm$^{2}$, with $s=1$, $%
\protect\theta _{1}=0.25^{\circ }$, $\protect\theta _{-1}=0.26^{\circ }$,
and $a_{1}^{2}=20$ TW/cm$^{2}$ (the respective peak intensity is $27.04$
TW/cm$^{2}$). The corresponding scaled parameters in Eqs. \eqref{NLSE2} and
\eqref{PARAMETERS2} are $M=8,$ $\protect\alpha =0.85$, and $\protect\eta %
=-0.02$, $b_{1}=0.87$. (b) The transverse intensity profile of the
quasi-stationary rotating azimuthon, measured at the propagation distance $%
z=114$ cm (the peak intensity of the profile is $24.25$ TW/cm$^{2}$).
Numerical evaluation of its angular velocity yields $\Omega =-1.75\pm 0.01$
deg/cm, with the sign corresponds to the clockwise direction. This value is
in close agreement with $\Omega =-1.758$ deg/cm given by Eq. (\protect\ref%
{Omega}) from the linear theory. (c) The nearly constant peak intensity and
NLL (the nonlinear-loss rate), defined as per Eq. \eqref{NLL} (black and red
continuous lines), vs. the propagation distance, $z$. Dashed lines represent
a non-rotating pattern for the same parameters, except that the two cone
angles are equal to $0.255^{\circ }$. (d) The corresponding static pattern ($%
\Omega =0$), not yet completely formed, at $z=100$ cm.}
\label{Fig1}
\end{figure}

\begin{figure}[tbp]
\includegraphics*[width=4.0cm]{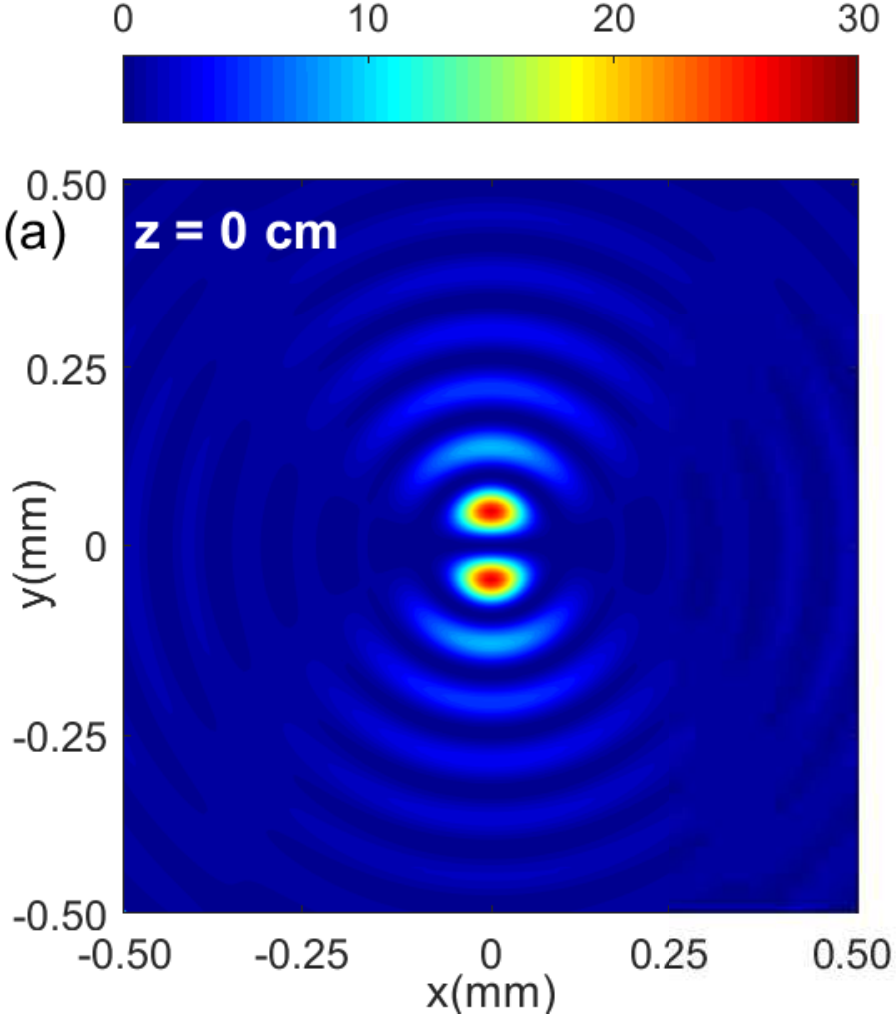} \includegraphics*[
width=4.0cm]{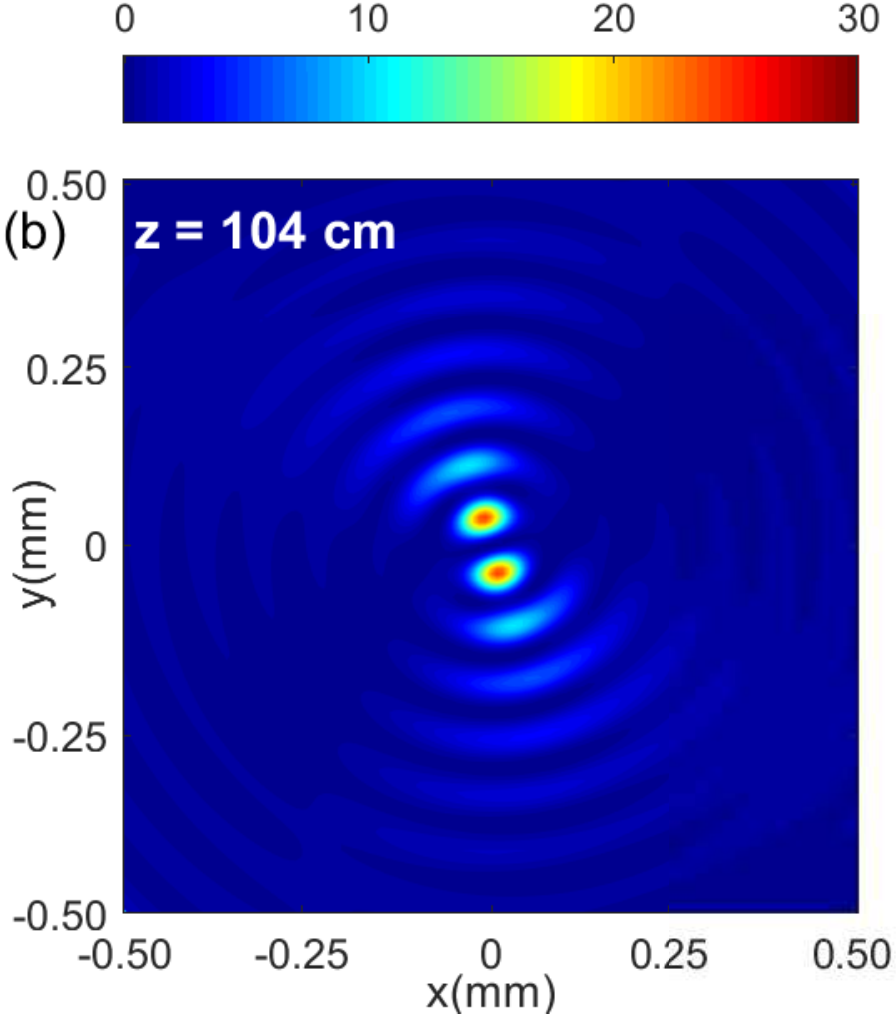} \includegraphics*[width=3.8cm]{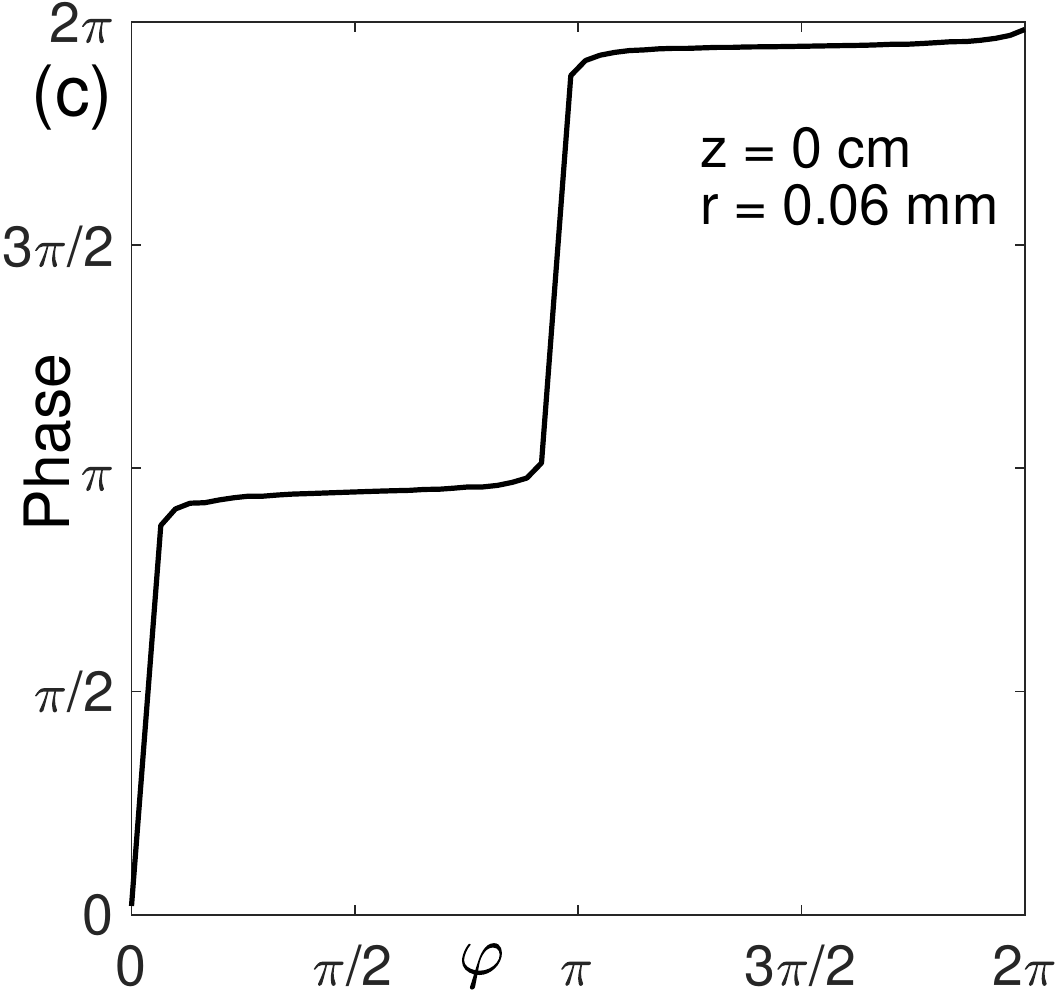} %
\includegraphics*[width=3.8cm]{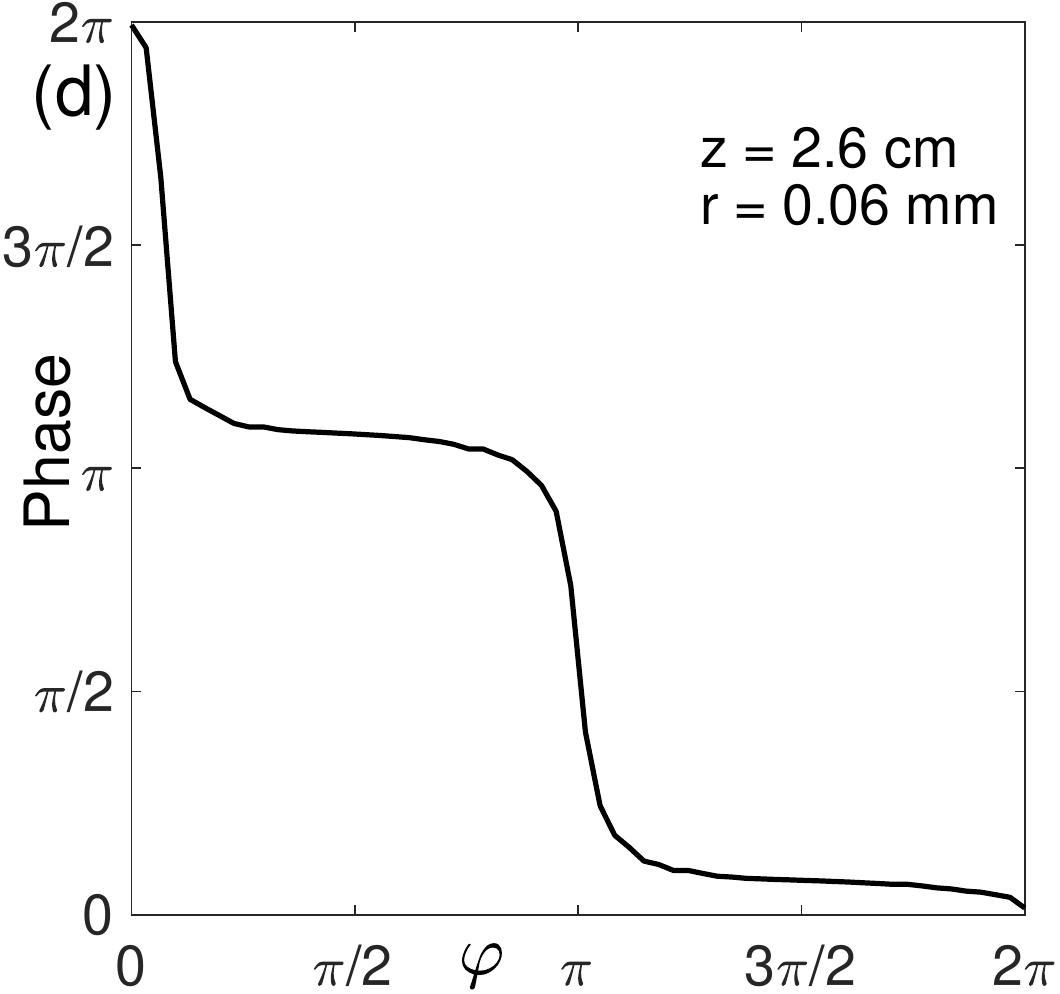}
\caption{Results for the setting with the same parameters as in Fig. \protect
\ref{Fig1}. (a) The transverse intensity profile, measured in TW/cm$^{2}$,
of the superposition of the Bessel beams in the input given by Eq.
\eqref{BESSEL-AZIMUTHON}, with $s=1$, $\protect\theta _{1}=0.26^{\circ }$, $%
\protect\theta _{-1}=0.24^{\circ }$, and $a_{1}^{2}=20$ TW/cm$^{2}$. The
corresponding scaled parameters in equations \eqref{NLSE2} and
\eqref{PARAMETERS2} are $M=8$, $\protect\alpha =0.88$, and $\protect\eta %
=0.04$, $b_{1}=0.87$, respectively. (b) The intensity profile of the
azimuthon produced by the propagation over distance $z=100$ cm. The
counterclockwise angular velocity is $\Omega =3.40\pm 0.08$ deg/cm from the
numerical simulations and $\Omega =3.447$ deg/cm from Eq. (\protect\ref%
{Omega}). See Supplemental Material at [URL of Fig2ab.avi] for the
excitation of the azimuthon from (a) to (b). Panels (c) and (d) display the
azimuthal phase profiles along a circle of small radius $r$ for the input
Bessel-beam superposition \eqref{BESSEL-AZIMUTHON}, and for the azimuthon
established by the propagation, respectively. See Supplemental Material at
[URL of Fig2cd.avi] for the transformation from (c) to (d) in the course of
the propagation, with the phase unwrapped in $[0,2\protect\pi ]$.}
\label{Fig2}
\end{figure}

In the context of filamentation, one can consider a more
accurate model in which $A$ is a function of time, and including material dispersion for pulsed beams, and plasma-defocusing effects.
Actually, the material dispersion may be neglected for propagation distances smaller than the
dispersion length, as confirmed by the theoretical and experimental
studies of the filamentation with Bessel beams \cite{JUKNA,XIU}. In terms of our
simulations displayed in Figs. \ref{Fig1} and \ref{Fig2}, dispersion effects in air would become
relevant at the dispersion length $z \approx 250$ m for typical pulses with
temporal duration $100$ fs. Therefore, the dispersion is negligible
for distances of $\approx 1$ m presented in this work. In particular,
it was theoretically and experimentally shown that the filamentation
dynamics of Bessel beams is correctly captured by the monochromatic model with $A$
multiplied by an invariant temporal pulse shape, as the key ingredients which
determine the filamentation are diffraction, Kerr self-focusing, and multiphoton absorption, while plasma
defocusing plays a secondary role \cite{JUKNA,XIU}.

In the absence of the Kerr and multiphoton absorption terms, Eq. \eqref{NLSE}
is satisfied by Bessel beams carrying vorticity with any integer topological
charge $s$. In the form explicitly representing the paraxial approximation,
they are $A(r,\varphi )\propto J_{s}(k\theta _{s}r)\exp (is\varphi )\exp
(i\delta _{s}z)$, where $J_{s}$ is the Bessel function of order $s$, $\theta
_{s}>0$ is the cone angle, and
\begin{equation}
\delta _{s}=-k\theta _{s}^{2}/2<0  \label{delta}
\end{equation}%
is the contribution to the propagation constant associated with the conical
geometry of the Bessel beam. In the linear regime, Eq. \eqref{NLSE} is
actually satisfied by any superposition of Bessel beams with arbitrary
topological charges, amplitudes and cone angles. In particular,
superpositions with different cone angles produce propagation-invariant
intensity patterns which rotate in the course of the propagation \cite%
{VASILYEU,ROP}. Rotatory polarization patterns in free space, emulating
optical activity of the effective medium, have also been demonstrated in
superpositions of orthogonally polarized Bessel beams with different cone
angles \cite{PORRASJOSAA}.

In nonlinear regimes, starting from Refs. \cite{JOHANISON} for the pure-Kerr
medium, and \cite{PORRASPRL,POLESANA} which included more general
nonlinearities and dissipative nonlinear terms, undistorted and inattenuated
propagation in the form of nonlinear Bessel beams was demonstrated
experimentally. A linear Bessel beam launched in the nonlinear medium
spontaneously transforms into an appropriately deformed beam. Subsequently,
similar results have been demonstrated for vortical Bessel beams, which too
were shown to transform into propagation-invariant nonlinear counterparts
\cite{PORRASJOSAB,JUKNA,XIU}. The existence of fully stable nonlinear Bessel
vortex beams in media with pure-cubic Kerr nonlinearity has been established
in Ref. \cite{PORRASPRA1}, where the stabilizing mechanism was provided by
nonlinear absorption. In a more recent study, arbitrary superpositions of
Bessel beams with different topological charges and amplitudes but identical
cone angles, launched into the medium with nonlinear absorption \cite%
{PORRASPRA2}, were shown to excite propagation-invariant
``dissipatons", which are nearly arbitrary structures composed of vortices and
bright spots where the power is continuously dissipated. In those studies,
stationarity propagation in media with nonlinear absorption was enabled by a
feeding mechanism provided by power influx from the reservoir with an
indefinitely large capacity, maintained by the slowly decaying quasi-linear
tails of nonlinear Bessel vortex beams and dissipatons. This mechanism is
not possible for strongly localized dissipative solitons with convergent
integral power, that should be maintained by intrinsic gain.

Searching for steady states in the rotating frame, we look for solutions to
Eq. (\ref{NLSE}) with input
\begin{gather}
A(r,\varphi ,z=0)=a_{s}\left[ J_{s}(k\theta _{s}r)\exp \left( is\varphi
\right) \right.   \notag \\
\left. +J_{-s}(k\theta _{-s}r)\exp \left( -is\varphi \right) \right] \,,
\label{BESSEL-AZIMUTHON}
\end{gather}%
composed of two Bessel beams with equal amplitudes $a_{s}$, opposite
vorticities $\pm s$, $s=1,2\dots $ and slightly different cone angles, $%
\theta _{\pm s}>0$, hence the respective shifts of the axial wave vectors,
\begin{equation}
\delta _{\pm s}=-k\theta _{\pm s}^{2}/2,  \label{delta+-}
\end{equation}%
are slightly different too. It is worthy to note that the solution of the
linearized version of Eq. (\ref{NLSE}) seeded by input (\ref%
{BESSEL-AZIMUTHON}), namely,
\begin{gather}
A(r,\varphi ,z)=a_{s}\left[ J_{s}(k\theta _{s}r)\exp \left( is\varphi
+i\delta _{s}z\right) \right.   \notag \\
+\left. J_{-s}(k\theta _{-s}r)\exp \left( -is\varphi +i\delta _{-s}z\right)
\right] \,,  \label{lin-Bessel}
\end{gather}%
features vorticity $s$ at its center ($r\rightarrow 0$) for $\theta
_{s}>\theta _{-s}$, and $-s$ for $\theta _{s}<\theta _{-s}$, with
counterclockwise and clockwise phase circulation, $\pm 2\pi s$,
respectively, along a circle of small radius $r$. With increasing radius,
the vorticity keeps switching between $s$ to $-s$. The angular velocity of
the rotation of the intensity pattern, $\left\vert A\left( r,\varphi
,z\right) \right\vert ^{2}$, corresponding to the linear solution in Eq. (%
\ref{lin-Bessel}) is \cite{VASILYEU,ROP}
\begin{equation}
\Omega =\frac{\delta _{-s}-\delta _{s}}{2s}=\frac{k}{4s}\left( \theta
_{s}^{2}-\theta _{-s}^{2}\right) ,  \label{Omega}
\end{equation}%
where Eq. (\ref{delta+-}) is used, which is counterclockwise (positive) for $%
\theta _{s}>\theta _{-s}$ and clockwise (negative) for $\theta _{s}<\theta
_{-s}$. Thus, the direction of the rotation of the intensity pattern
coincides with the direction of the azimuthal gradient of the phase close to
the beam's center in these linear superpositions of Bessel beams.

Figure \ref{Fig1}(a) displays an example of the nonlinear evolution produced
by initial condition in Eq. \eqref{BESSEL-AZIMUTHON} with $s=1$ and slightly
different cone angles. The simulations of Eq. (\ref{NLSE}) were performed %
(as mentioned above, in both Cartesian and polar
coordinates) with parameters corresponding to the propagation of
high-intensity light in air, in the regime for which the Kerr self-focusing
and nonlinear absorption are significant (see the caption to the figure for
details). The numerical solution demonstrates that the propagating beam
quickly attains a steady rotatory state, featuring compressed lobes as a
result of the Kerr self-focusing, as seen in Fig. \ref{Fig1}(b). The solid
black curve in Fig. \ref{Fig1}(c) shows that the peak intensity approaches a
constant value in the course of the propagation, confirming the stationarity
of the rotating pattern. Another manifestation of the steady propagation is
the fact that the rate of the nonlinear loss (NLL) of the power,
\begin{equation}
\mathrm{NLL}=2\pi \beta ^{(M)}\int_{0}^{\infty }|A(r)|^{2M}rdr\,,
\label{NLL}
\end{equation}%
also attains a nearly constant value, as shown by the red solid curve in
Fig. \ref{Fig1}(c). Thus, the rotating structure is a genuine dissipative
azimuthon.

\begin{figure}[b]
\includegraphics*[width=8.5cm]{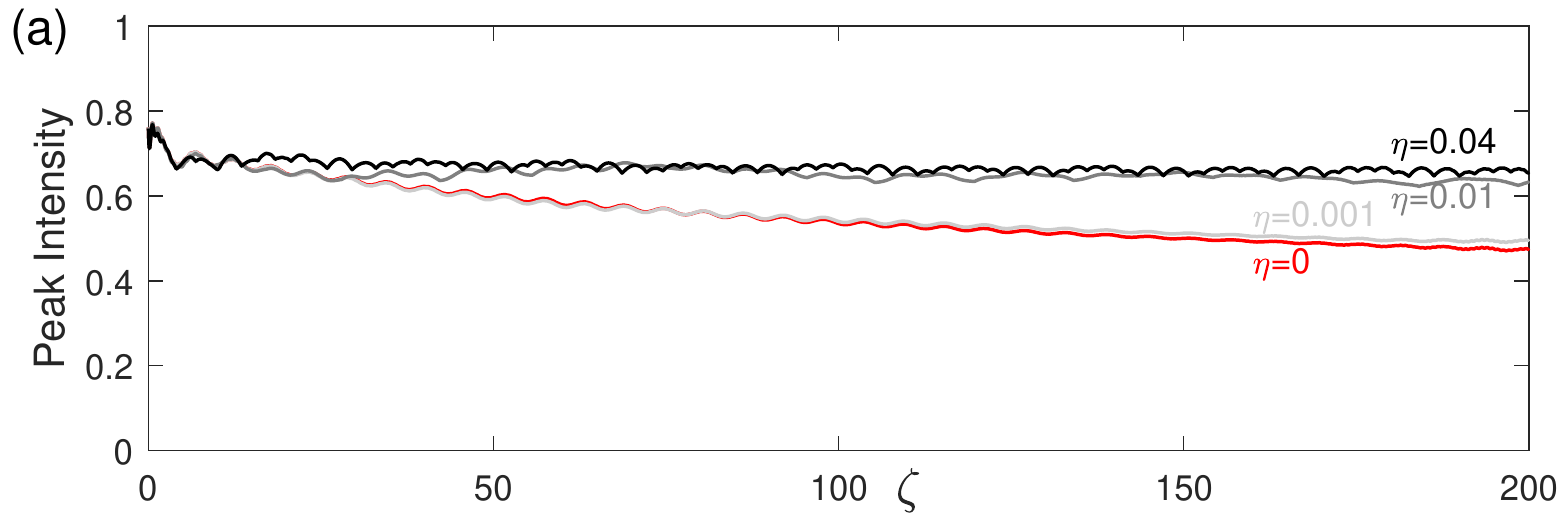} \includegraphics*[
width=4.25cm]{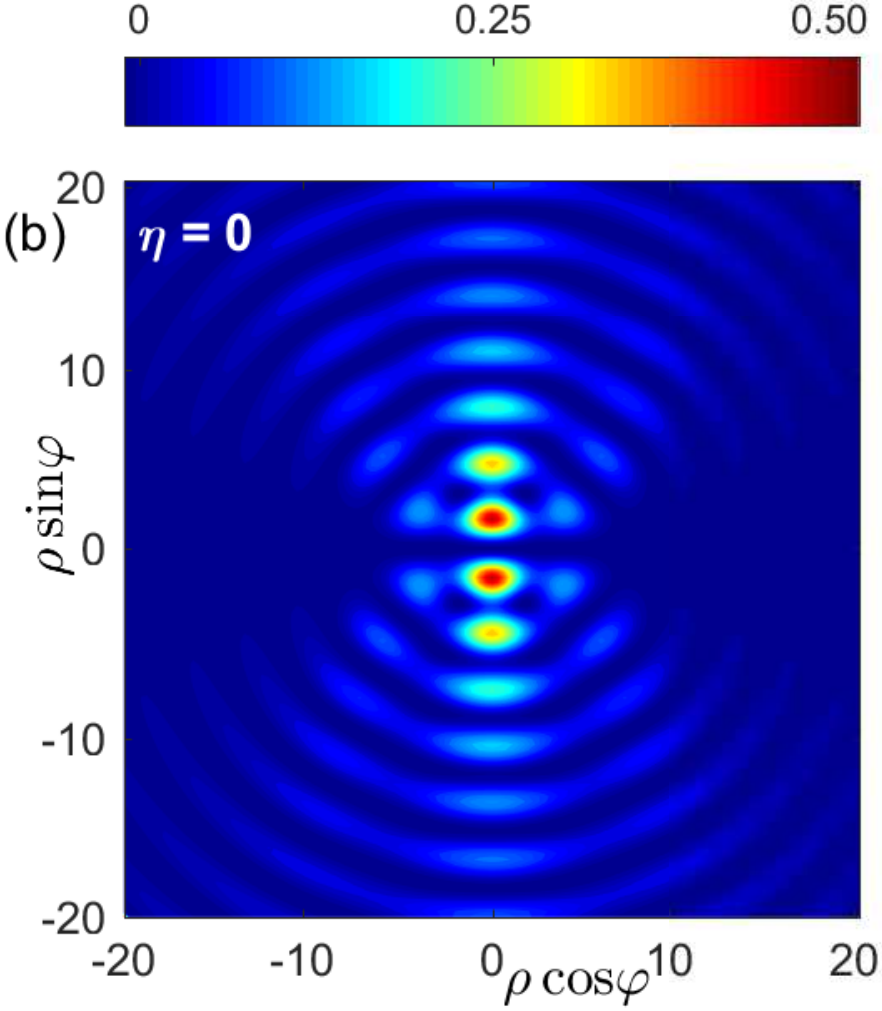} \includegraphics*[width=4.25cm]{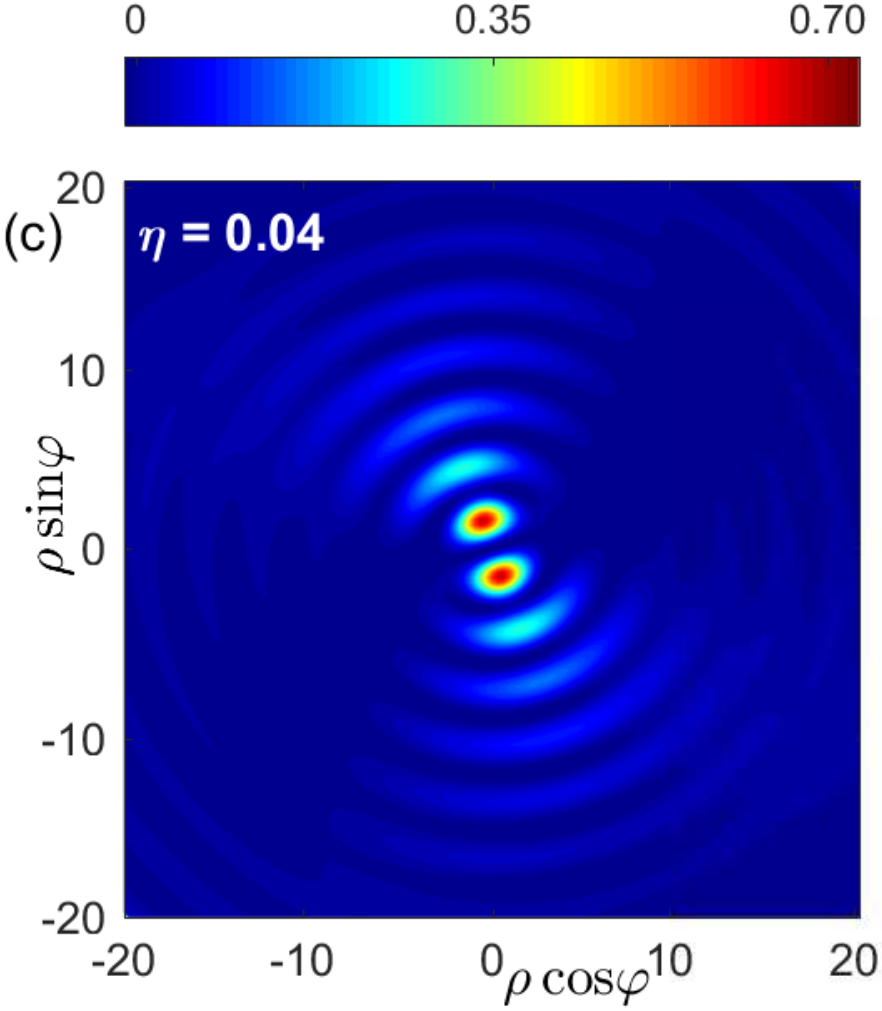}
\caption{(a) Peak intensities as functions of the propagation distance for
the dissipative azimuthons formed, at values of parameters $M=4$, $\protect%
\alpha =1$, starting with the Bessel-beam superposition input given by Eq. (%
\protect\ref{BESSEL-AZIMUTHON3}), with $s=1$, $b_{1}=0.75$ and increasing
values of scaled mismatch, $\protect\eta =0,0.001,0.01$ and $0.04$, see Eq. (%
\protect\ref{PARAMETERS2}). The characteristic relaxation distance for $%
\protect\eta =0$ is $\protect\zeta _{\mathrm{rel}}\simeq 70$, yielding $|%
\protect\eta |_{\mathrm{\min }}\simeq 0.04$, see Eq. (\protect\ref%
{relaxation}). (b) The intensity pattern of the non-rotating dissipaton
formed with $\protect\eta =0$ and $\protect\varpi =0$. (c) The same as in
(b) but for the rotating dissipative azimuthon with $\protect\eta =0.04$ and
counterclockwise scaled angular velocity, $\protect\varpi =0.081\pm 0.003$
as obtained from the simulations, and $\protect\varpi =0.080$, as produced
by Eq. (\protect\ref{omega}). The latter pattern forms after propagating a
much shorter distance. }
\label{Fig3}
\end{figure}

\begin{figure}[t]
\begin{center}
\subfigure[ ]{\includegraphics*[width=4.0cm]{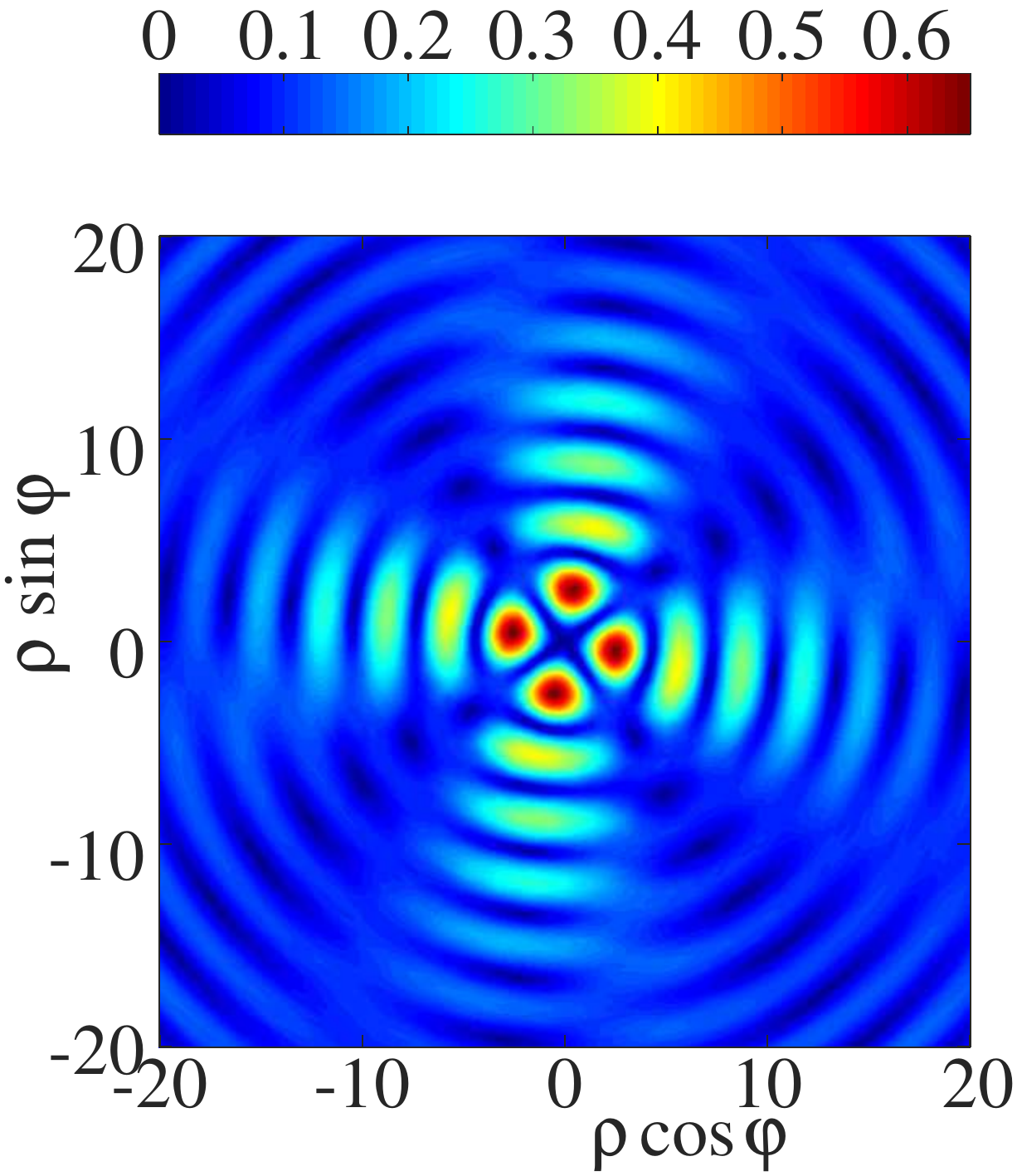}}
\subfigure[
]{\includegraphics*[width=4.0cm]{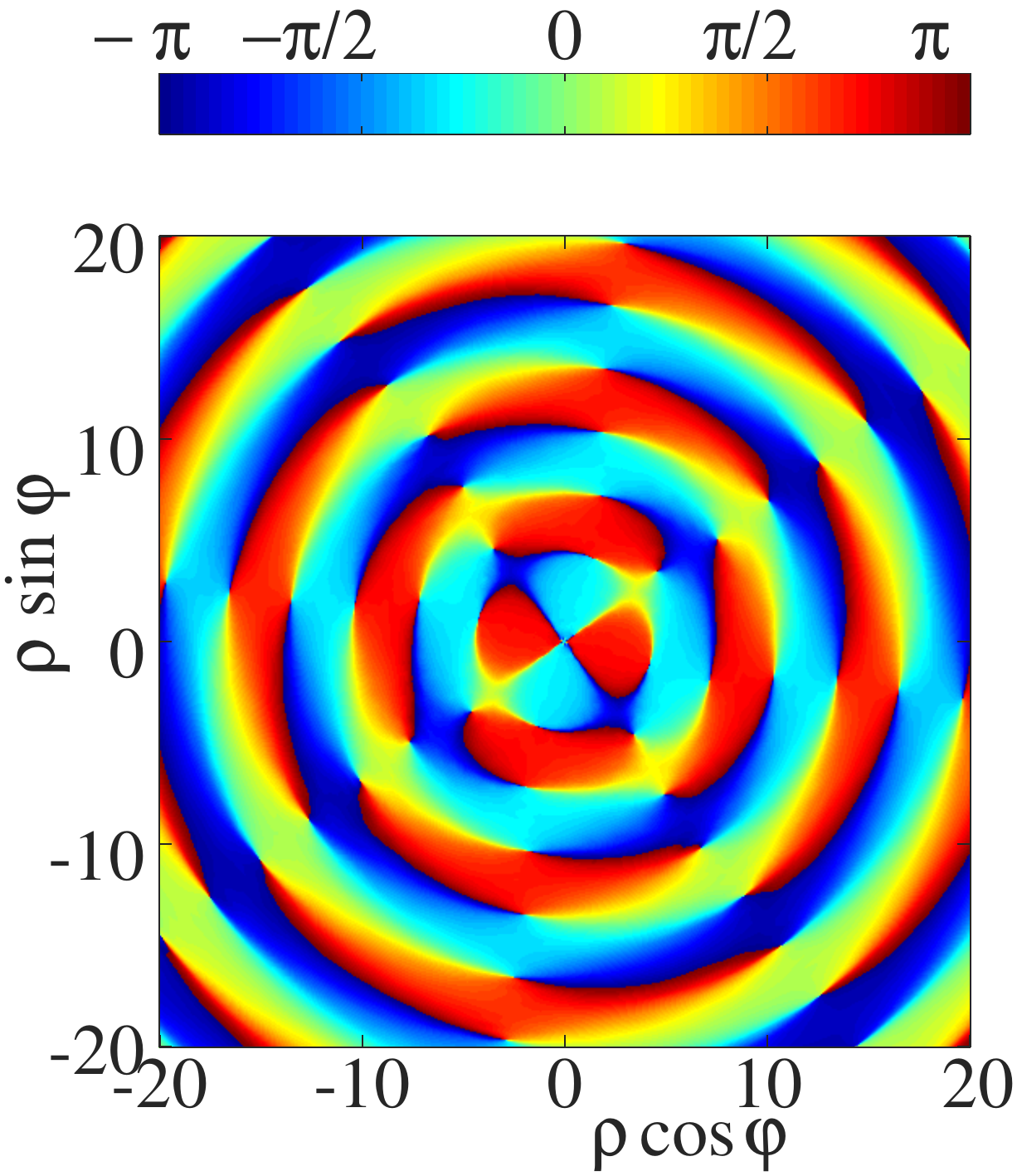}}
\subfigure[
]{\includegraphics*[width=4.0cm]{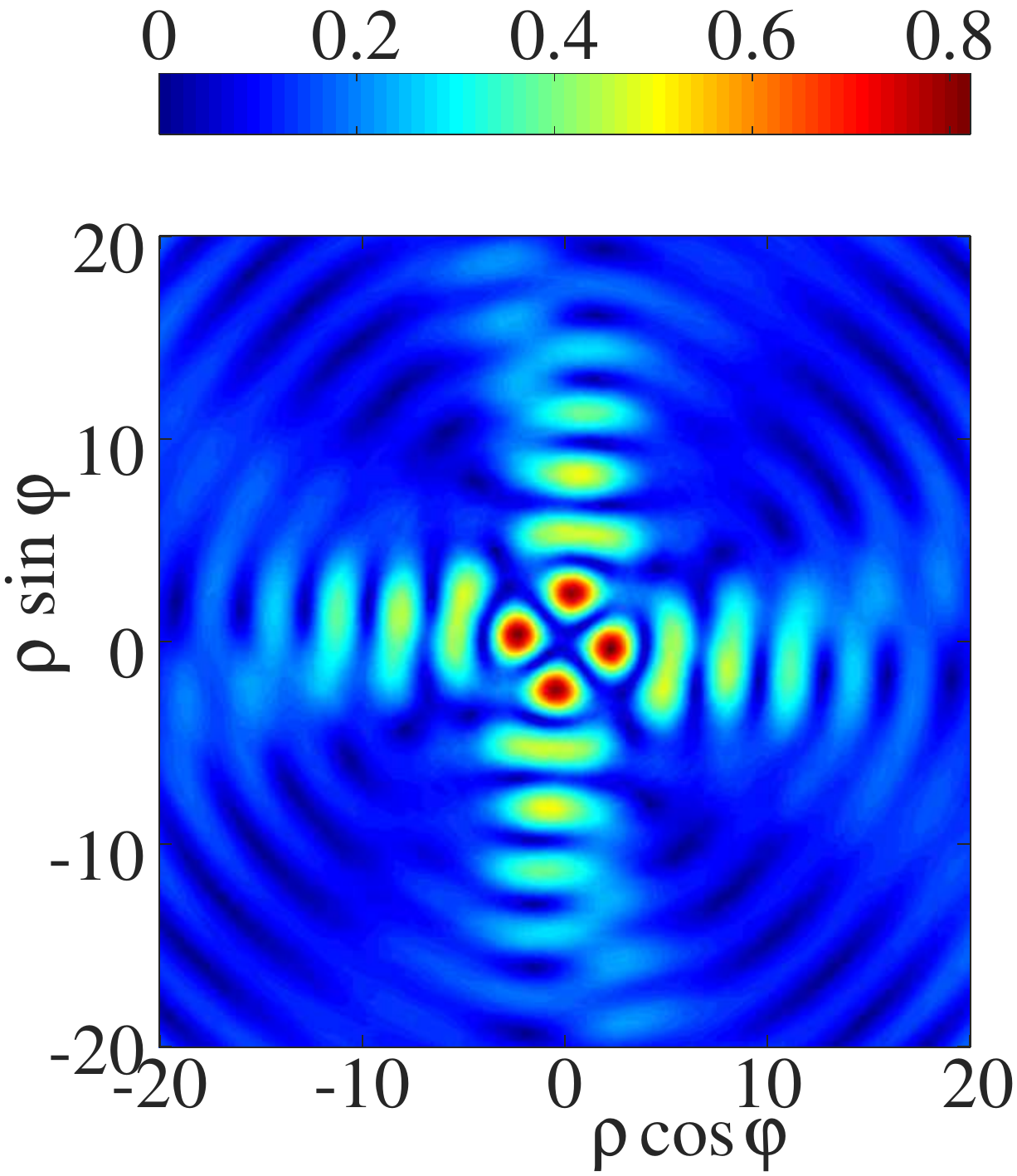}}
\subfigure[
]{\includegraphics*[width=4.0cm]{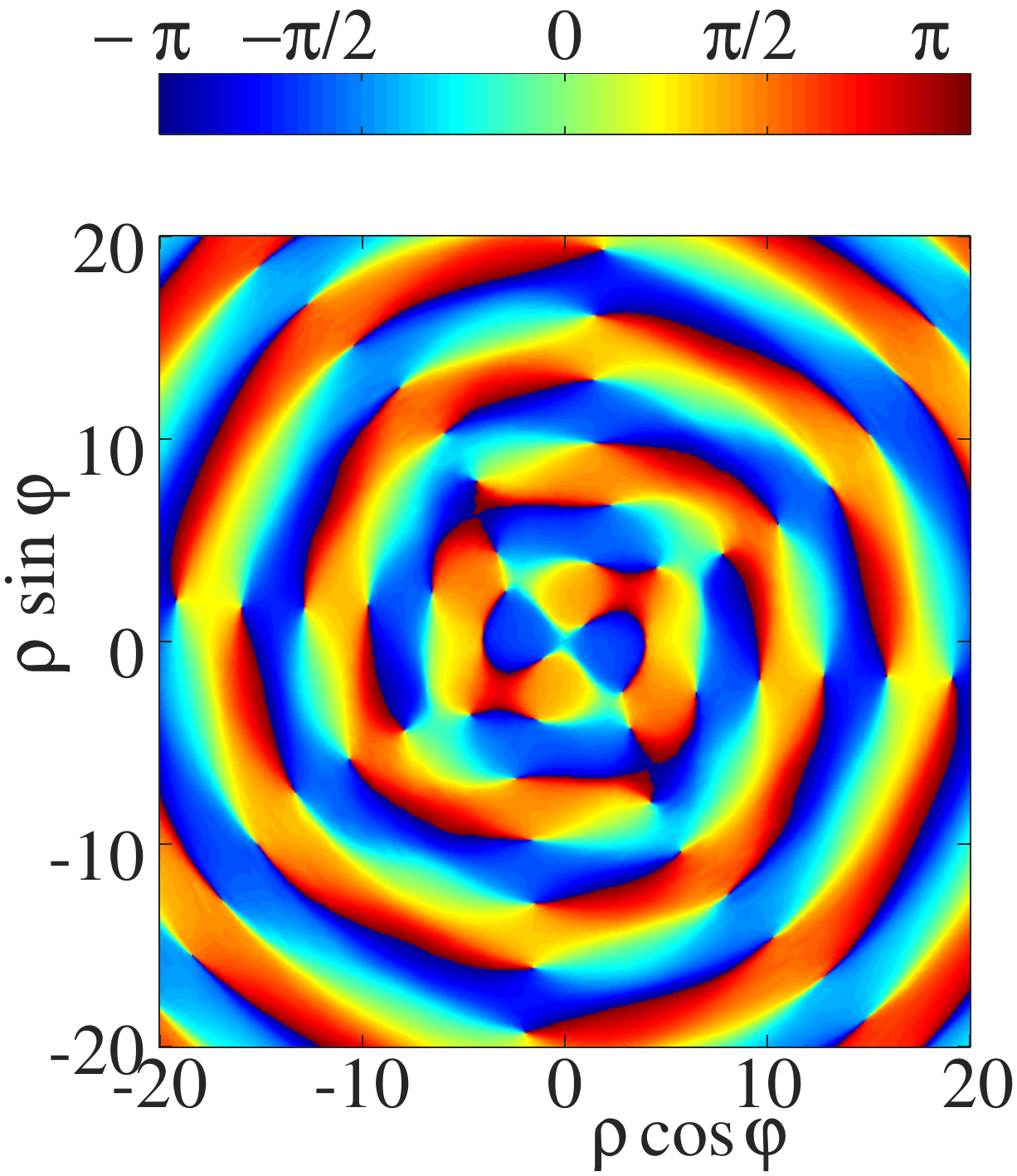}}
\end{center}
\caption{Stable dissipative azimuthons, spontaneously formed, in the setting
with $M=4$ and Kerr coefficient $\protect\alpha =1.8$, from input (\protect
\ref{BESSEL-AZIMUTHON}) with $s=2$ and $\protect\eta =0.04$. (a) The
amplitude and (b) phase profiles at $\protect\zeta =270$, obtained starting
with the scaled Bessel amplitude $|b_{s}|=0.6$. The angular velocity is $%
\protect\varpi \simeq 0.0403\pm 0.0002$, while $2\protect\eta /s=0.04$, see
Supplemental Material [URL of Fig4a.avi] for the excitation of the azimuthon
in (a) from the input Bessel-beam superposition. Panels (c) and (d) display
the same as (a) and (b), but for $\protect\alpha =2$ and $|b_{s}|=0.8$, the
angular velocity beubg $\protect\varpi \simeq 0.0403\pm 0.0004$.}
\label{Fig4}
\end{figure}

\begin{figure}[t]
\begin{center}
\subfigure[ ]{\includegraphics*[width=4.0cm]{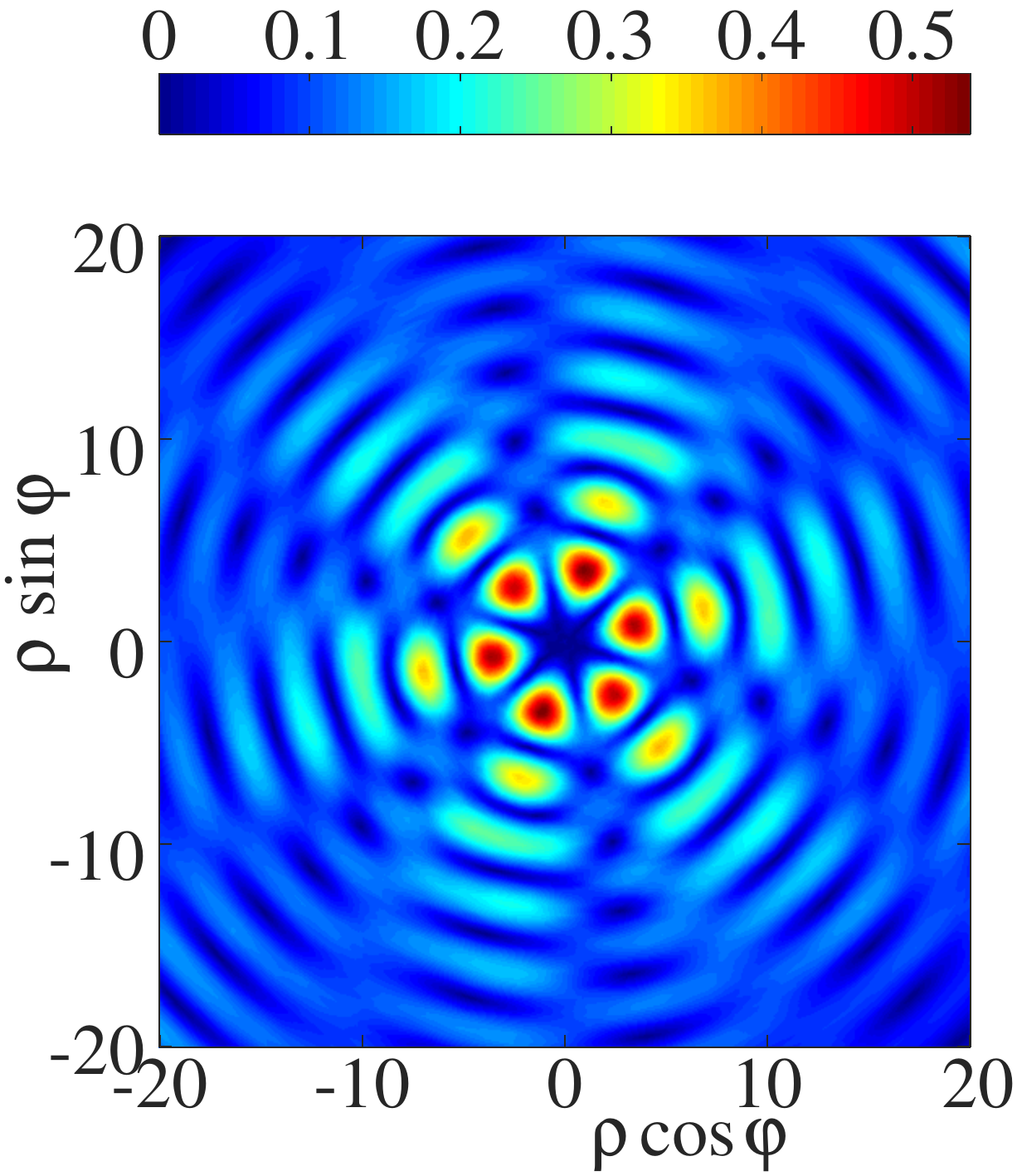}}
\subfigure[
]{\includegraphics*[width=4.0cm]{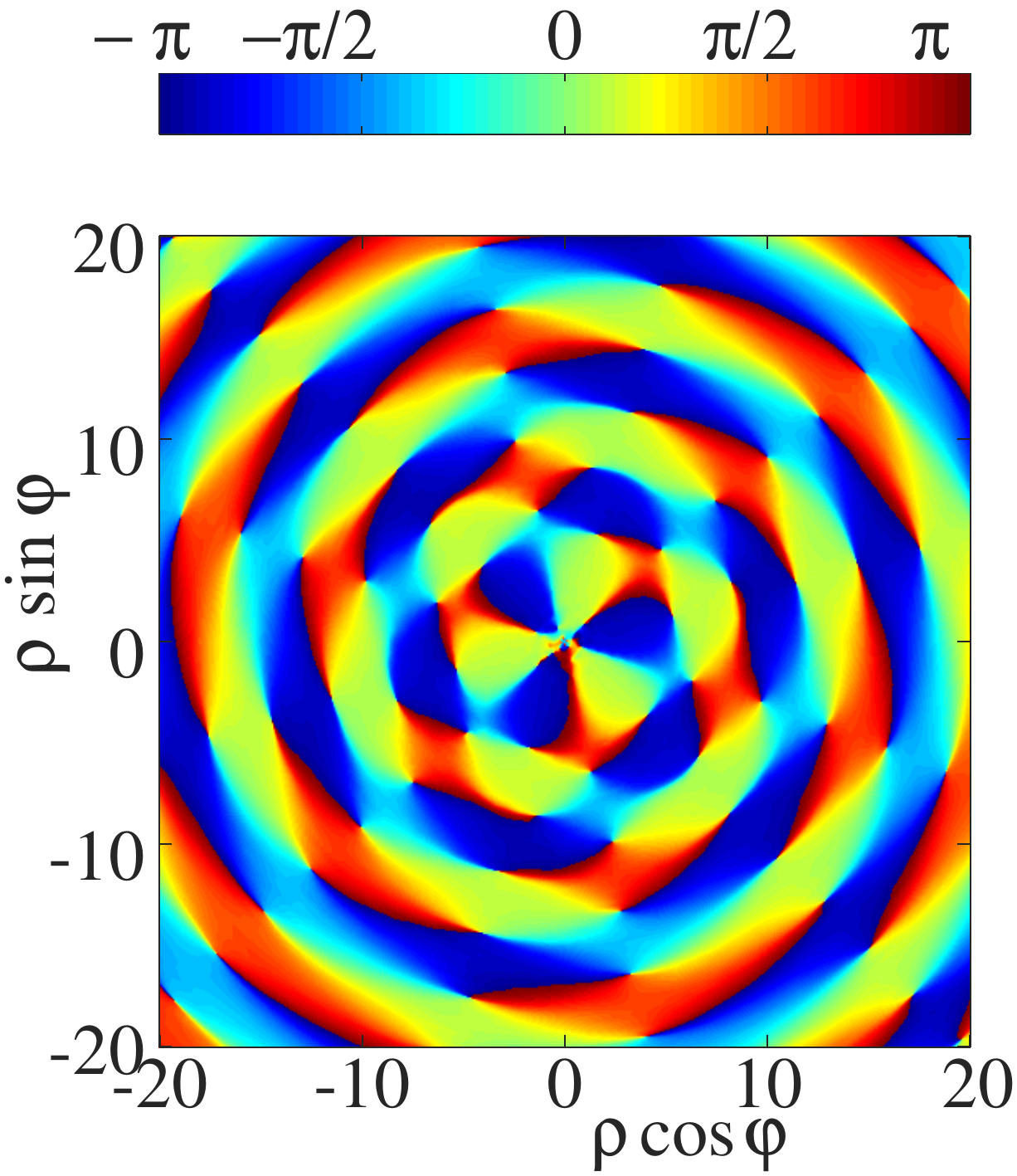}}
\subfigure[
]{\includegraphics*[width=4.0cm]{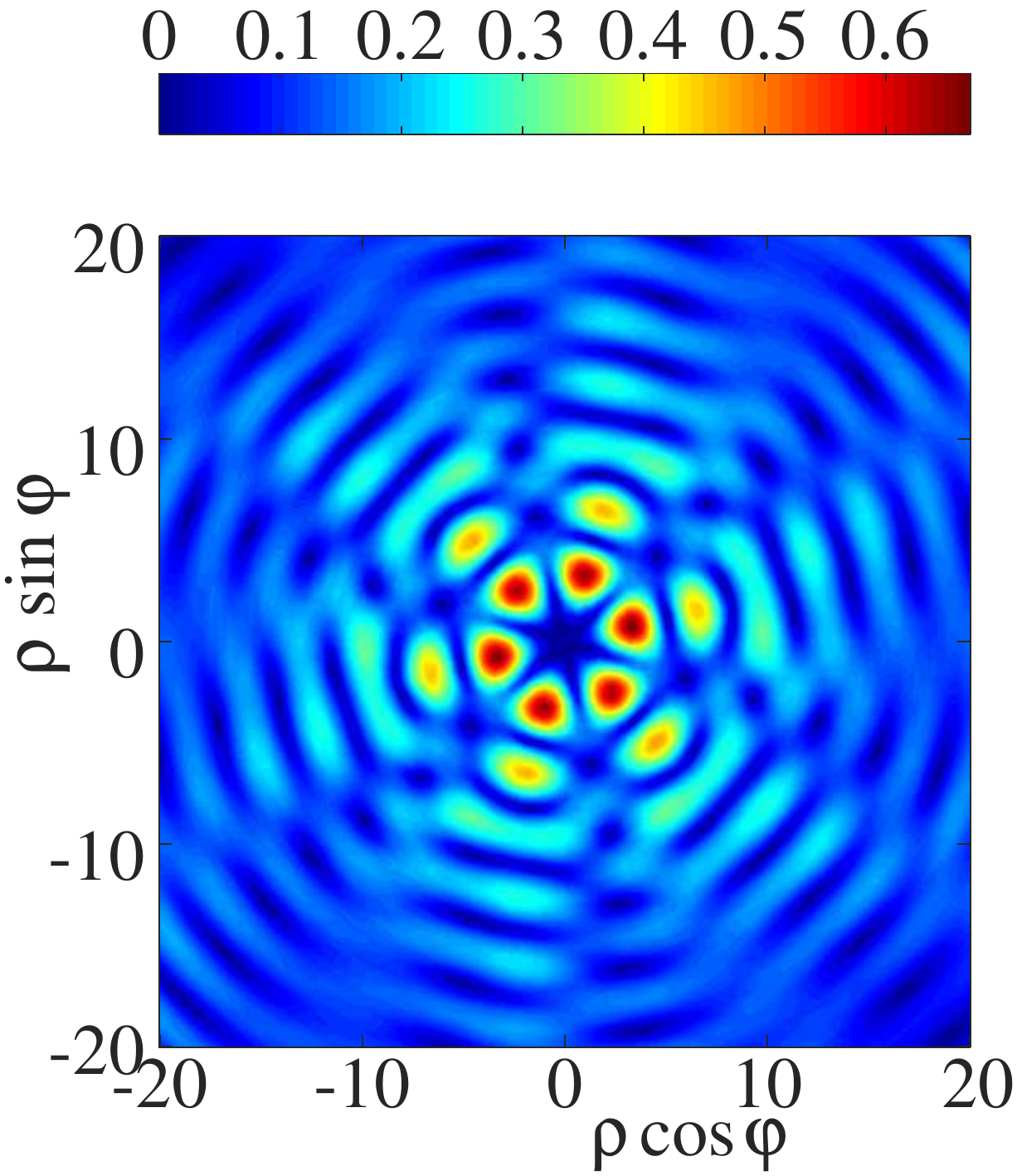}}
\subfigure[
]{\includegraphics*[width=4.0cm]{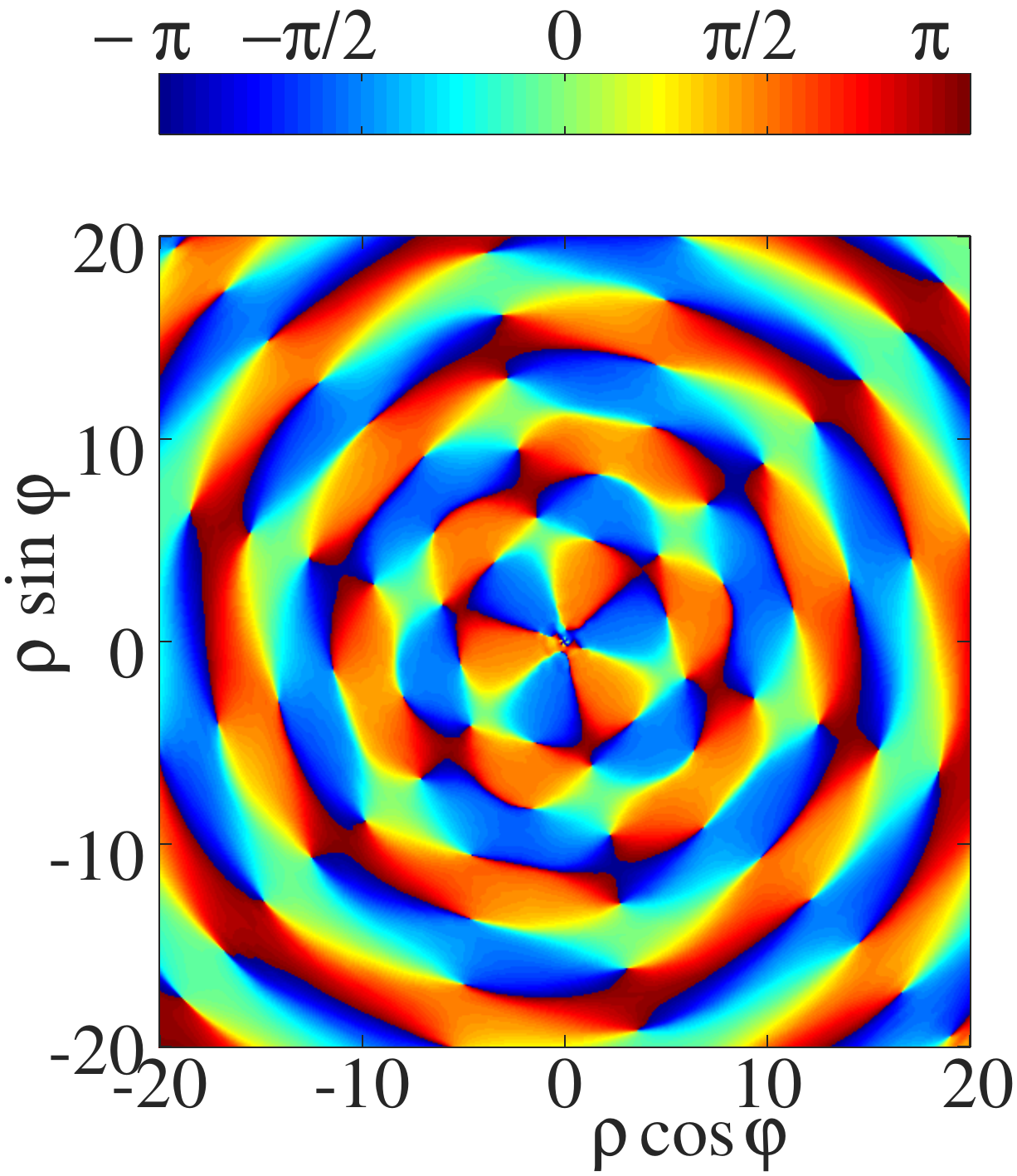}}
\end{center}
\caption{Stable dissipative azimuthons established in the setting with $M=4$
and Kerr coefficient $\protect\alpha =2.2$, starting from input (\protect\ref%
{BESSEL-AZIMUTHON}) with $s=3$ and $\protect\eta =0.04$. (a) The amplitude
and (b) phase structure at $\protect\zeta =225$, obtained starting with the
scaled Bessel amplitude $|b_{s}|=0.6$. The angular velocity is $\protect%
\varpi \simeq 0.0267\pm 0.0003$, while $2\protect\eta /s\simeq 0.0267$.
Panels (c) and (d) display the same as (a) and (b), but for $\protect\alpha %
=1.9$ and $|b_{s}|=0.8$, the angular velocity being $\protect\varpi \simeq
0.0268\pm 0.0002$.}
\label{Fig5}
\end{figure}

As shown in Figs. \ref{Fig1}-\ref{Fig5} for different values of $s$,
intensity patterns of dissipative azimuthons preserve the $2s$-fold
rotational symmetry of the input configuration given by Eq. (\ref%
{BESSEL-AZIMUTHON}). Thus, the number of ``hot spots" in the
rotating azimuthons, $N=2s$, is only determined by magnitude $s$ of the
topological charge in the two beams which build the input. In this respect,
dissipative azimuthons differ from their conservative counterparts and
soliton clusters, for which the number of hot spots and the vorticity at the
center are independent integers \cite{DESYATNIKOV2,MINO}. The latter
property of the conservative model is explained by the fact that the number
of intensity maxima in the circular pattern is determined by the
modulational instability of the axially uniform state.

Another manifestation of the robustness of the dissipative azimuthons is
that they preserve the angular velocity imposed by the superposition of the
two vortex Bessel beams in input (\ref{BESSEL-AZIMUTHON}), \textit{viz}., $%
\Omega =(\delta _{-s}-\delta _{s})/2s$; hence the rotation period is $z_{%
\mathrm{rotation}}=4\pi s/|\delta _{-s}-\delta _{s}|$, and, given the $2s$%
-fold rotational symmetry of the rotating pattern, it periodically repeats
itself after passing distance $2\pi /|\delta _{-s}-\delta _{s}|$,
independent of the topological charge, $s$. The preservation of the angular
velocity is explained by the fact that the dissipative azimuthon is
surrounded by the quasi-linear asymptotic field (providing the
above-mentioned power reservoir) that continues to rotate as the input
linear superposition does, i.e., with angular velocity (\ref{Omega}). Since
the whole structure is stationary in the rotating frame, the inner nonlinear
region rotates synchronously with the small-amplitude periphery.

A general feature of dissipative azimuthons that makes them substantially
different from the linear Bessel superposition in Eq. (\ref{lin-Bessel}) is
that the direction of rotation of the intensity pattern is \emph{opposite}
to the direction of the azimuthal phase gradient close to the pattern's
pivot. The reason is that the topological charge of the vortex at the center
reverses its sign in the course of the azimuthon formation from the initial
Bessel beam superposition, as shown in Fig. \ref{Fig2}. The intensity
pattern of the input Bessel-beam superposition in Fig. \ref{Fig2}(a) has $%
\theta _{s}>\theta _{-s}$, hence the input vorticity at the center is $s=1$,
corresponding to the counterclockwise phase increase by $2\pi s$ around the
origin, as shown by the azimuthal phase profile in Fig. \ref{Fig2}(c).
Accordingly, the intensity pattern in the azimuthon established by the
evolution of the intensity pattern, which is displayed in Fig. \ref{Fig2}%
(b), rotates counterclockwise too. However, the vorticity at the center
inverts in the course of the evolution to $-s=-1$, corresponding to the
clockwise phase circulation $2\pi s$ around the origin, as seen in Fig. \ref%
{Fig2}(d). Details of the transformation of the azimuthal phase profile from
vorticity $s=1$ in Fig. \ref{Fig2}(c) to $-s=-1$ in Fig. \ref{Fig2}(d) in
the course of the formation of the azimuthon are shown in Supplemental
Material [URL of Fig2cd.avi]. The vorticity inversion affects the entire
central zone of the azimuthon, including the hot spots. At larger radii the
vorticity oscillates, and in the peripheral zone the periodic alternations
coincide with those of the input Bessel-beam superposition.

It is relevant to note that a possibility of the dynamical inversion of the
sign of the topological charge of an optical vortex in a conservative
medium, which interacts with a material lattice structure, was previously
predicted theoretically \cite{Kivshar} and demonstrated experimentally \cite%
{Zhigang}. In the present setting, the rotating intensity pattern emerging
in the azimuthon mode may play a role of such a lattice. Indeed, the
rotation of a layer with radius $R$ at angular velocity $\Omega $ tends to
add the term generated by the Galilean transform, $kR^{2}\Omega \left(
\varphi -\Omega z/2\right) $, to the phase of the wave field (subject to the
periodicity constraint, $kR^{2}\Omega =m$, with integer $m$). If, on the
other hand, vorticity phase $\pm s\varphi $ dominates at $r$ small enough,
the minimization of the gradient term in the Hamiltonian of the conservative
part of Eq. (\ref{NLSE}), with density $\left( 2k\right) ^{-1}\left\vert
\nabla _{\perp }A\right\vert ^{2}$, suggests to choose mutual signs of $%
\Omega $ and $s$ which help to cancel different contributions to the phase.

Thus, dissipative azimuthons clearly exhibit a mixed linear-nonlinear
structure: The stationary intensity pattern, including the quasi-linear
periphery, rotating with angular velocity given by Eq. (\ref{Omega}), which
is imposed by input (\ref{BESSEL-AZIMUTHON}), and a restructured nonlinear
core, including the circular chain of hot spots, whose vorticity is inverted
with respect to the rotation direction, as in soliton clusters \cite%
{DESYATNIKOV1,DESYATNIKOV2}.

For a more comprehensive study of properties of the dissipative azimuthons
[in particular, their (in)stability], we introduce
\begin{equation}
\delta \equiv \frac{1}{2}\left( \delta _{s}+\delta _{-s}\right) ,\qquad
\Delta \delta \equiv \frac{1}{2}\left( \delta _{s}-\delta _{-s}\right) \,,
\label{PARAMETERS}
\end{equation}%
where $\delta _{\pm s}$ is defined as per Eq. (\ref{delta}), and the scaled
radius and propagation distance,
\begin{equation}
\rho \equiv \sqrt{2k|\delta |}r,\quad \zeta \equiv |\delta |z,\quad \tilde{A}%
\equiv \left( \frac{\beta ^{(M)}}{2|\delta |}\right) ^{\frac{1}{2M-2}}A\,.
\label{SCALING}
\end{equation}%
These rescalings transform NLSE \eqref{NLSE} into
\begin{equation}
\partial _{\zeta }\tilde{A}=i\nabla _{\perp }^{2}\tilde{A}+i\alpha |\tilde{A}%
|^{2}\tilde{A}-|\tilde{A}|^{2M-2}\tilde{A}\,,  \label{NLSE2}
\end{equation}%
where now $\nabla _{\perp }^{2}=\partial _{\rho }^{2}+(1/\rho )\partial
_{\rho }+(1/\rho ^{2})\partial _{\varphi }^{2}$, and
\begin{equation}
\alpha \equiv \left( \frac{2|\delta |}{\beta ^{(M)}}\right) ^{1/(M-1)}\frac{%
kn_{2}}{n|\delta |}\,.  \label{alpha}
\end{equation}%
In this notation, input (\ref{BESSEL-AZIMUTHON}) takes the form of
\begin{gather}
A(\rho ,\varphi ,0)=b_{s}\left[ J_{s}\left( \sqrt{1+\frac{\Delta \delta }{%
\delta }}\,\rho \right) e^{is\varphi }\right.   \notag \\
+\left. J_{-s}\left( \sqrt{1\!-\!\frac{\Delta \delta }{\delta }}\,\rho
\right) e^{-is\varphi }\right] \,,
\end{gather}%
which, given the smallness of $|\Delta \delta /\delta |$ for close values of
the cone angles, may be approximated by
\begin{equation}
A(\rho ,\varphi ,0)=b_{s}\left[ J_{s}\left( (1+\eta )\rho \right)
e^{is\varphi }+J_{-s}\left( (1-\eta )\rho \right) e^{-is\varphi }\right] ,
\label{BESSEL-AZIMUTHON3}
\end{equation}%
where we set $\sqrt{1\pm \Delta \delta /\delta }\simeq 1\pm \eta $, and
\begin{equation}
\eta \equiv \frac{\Delta \delta }{2\delta },\quad b_{s}\equiv \left( \frac{%
\beta ^{(M)}}{2|\delta |}\right) ^{\frac{1}{2M-2}}a_{s}\,.
\label{PARAMETERS2}
\end{equation}

Properties of dissipative azimuthons, similar to what was previously
reported for static dissipations and nonlinear Bessel vortex beams, are
essentially the same, regardless of the multiphoton absorption order $M$ in
Eq. \ref{BESSEL-AZIMUTHON}, therefore we henceforth fix $M=4$ (note, in
particular, that $M=4$ takes place in water at $527$ nm).
The dissipative azimuthon is then determined by vorticity $s$, the strength
of the Kerr nonlinearity relative to the nonlinear absorption, $\alpha $,
the amplitude parameter, $b_{s}$, and the radial mismatch, $\eta $ of the
input Bessel beams, see Eqs. (\ref{alpha}) and (\ref{PARAMETERS2}). Positive
and negative $\eta $ correspond, respectively, to $\theta _{s}>\theta _{-s}$
and $\theta _{s}<\theta _{-s}$, i.e., positive (counterclockwise) and
negative (clockwise) angular velocity. In the scaled notation, the angular
velocity is
\begin{equation}
\varpi =\frac{2\eta }{s},  \label{omega}
\end{equation}%
and the rotation period is $\zeta _{\mathrm{rotation}}=\pi s/|\eta |$. The
above-mentioned inversion of the vorticity implies that the respective
topological charges are $-s$ and $s$ for $\eta >0$ and $\eta <0$,
respectively. STOP

\section{The gyroscopic effect in dissipative azimuthons}

In comparison to nonlinear Bessel beams and static dissipatons \cite%
{PORRASJOSAB,PORRASPRA2}, it is seen in Figs. \ref{Fig1} and \ref{Fig3} that
the rotation accelerates the formation of the dissipative patterns. The
dashed curves in Fig. \ref{Fig1}(c) show the peak intensity and nonlinear
power loss rate as functions of the propagation distance, and Fig. \ref{Fig1}%
(d) displays the static intensity pattern established, after some
propagation distance, in the nonlinear medium, when the cone angles of the
input Bessel beams are made equal to the mean value of the two slightly
different angles of the original input. It is seen that the profile with the
equal conicities is not yet formed because it still continues to decrease
its amplitude, in comparison to the original one that propagates without
attenuation. The non-rotating pattern requires a much longer propagation
distance to form than its steadily rotating counterpart, whose formation is
promoted by what may be called a \textit{gyroscopic effect}. Namely, if the
static pattern is created by input (\ref{BESSEL-AZIMUTHON3}) with $\eta =0$,
the slow relaxation towards the static state is characterized by a certain
relaxation length, $\zeta _{\mathrm{rel}}$. On the other hand, if slight
difference of conicities of the two Bessel beams in the input promotes the
formation of a steadily rotating state, characterized by a rotation period $%
\zeta _{\mathrm{rotation}}=\pi s/|\eta |$. The rotation is expected to
eclipse the slow relaxation provided that $\zeta _{\mathrm{rotation}%
}\lesssim \zeta _{\mathrm{rel}}$, i.e., for
\begin{equation}
|\eta |\gtrsim |\eta |_{\mathrm{\min }}=\pi s/\zeta _{\mathrm{rel}}.
\label{relaxation}
\end{equation}%
In the example displayed Fig. \ref{Fig3} with $s=1$, $b_{1}=0.75$ and $M=4$,
$\alpha =1$, the characteristic relaxation distance, $\zeta _{\mathrm{rel}%
}\sim 70$, of the non-rotating pattern predicts $|\eta |_{\mathrm{\min }%
}\simeq 0.04$. As seen in Fig. \ref{Fig3}(a), the long relaxation stage,
following the short initial stage of compression under the combined action
of the self-focusing and absorption, is indeed eliminated at $\eta \gtrsim
|\eta |_{\min }$. Eventually, the snapshot of the intensity pattern in the
rotating state is quite similar to that of the non-rotating one, cf. Figs. %
\ref{Fig3}(b) and (c), but the rotating pattern forms much faster.

\section{Stability of the dissipative azimuthons}

\subsection{Stability limits}

Similar to nonlinear Bessel vortex beams \cite{PORRASPRA1} and dissipatons
\cite{PORRASPRA2}, dissipative azimuthons remain stable for sufficiently low
values of the normalized Kerr coefficient $\alpha $ [see Eq. (\ref{alpha})],
and they become unstable above a threshold value, $\alpha >\alpha _{\mathrm{%
th}}$, which is significantly higher than the instability threshold for
non-rotating states \cite{PORRASPRA1,PORRASPRA2}, implying that the rotation
enhances the stability. By means of systematic simulations, we have
identified the threshold as a function of $\alpha $ for several values of
vorticity $s$, using input (\ref{BESSEL-AZIMUTHON3}).

Given the large number of parameters, in the numerical simulations we fixed $%
|b_{s}|=0.6$ and $|b_{s}|=0.8$ [see Eq. (\ref{PARAMETERS2})], which
represent typical intensities at which the Kerr nonlinearity and multiphoton
absorption are substantial in fused silica, and $\eta =0.04$, which implies
relative variations of the cone angle $\simeq 10\%$.

For $s=1$, the threshold is found to be $\left( \alpha _{\mathrm{th}}\right)
_{s=1}=2.25\pm 0.05$ for $|b_{s}|=0.6$ and $|b_{s}|=0.8$. The dissipative
azimuthons with a higher vorticity are somewhat less stable, featuring $%
\left( \alpha _{\mathrm{th}}\right) _{s=2}=1.95\pm 0.05$ for $|b_{s}|=0.6$,
and $\left( \alpha _{\mathrm{th}}\right) _{s=2}=2.15\pm 0.05$ for $%
|b_{s}|=0.8$. Further, for $s=3$, it was found that $\left( \alpha _{\mathrm{%
th}}\right) _{s=3}=2.25\pm 0.05$ for $|b_{s}|=0.6$, and $\left( \alpha _{%
\mathrm{th}}\right) _{s=3}=1.95\pm 0.05$ for $|b_{s}|=0.8$. The difference
between the cases of $s=2$ and $3$ plausibly originates from their differing
symmetries. These instability thresholds are about twice as large as typical
values $\alpha _{\mathrm{th}}\simeq 1$ for nonlinear Bessel beams and
non-rotating dissipatons found in Refs. \cite{PORRASPRA1} and \cite%
{PORRASPRA2}, respectively, which indicates the enhanced robustness provided
by the rotation. Figures \ref{Fig4} and \ref{Fig5} show intensity and phase
patterns of the rotating dissipative azimuthons with $s=2$ and $3$, and $%
\alpha =1$, which places them well within the stability region.

\begin{figure}[b]
\begin{center}
\subfigure[]{\includegraphics*[width=4.0cm]{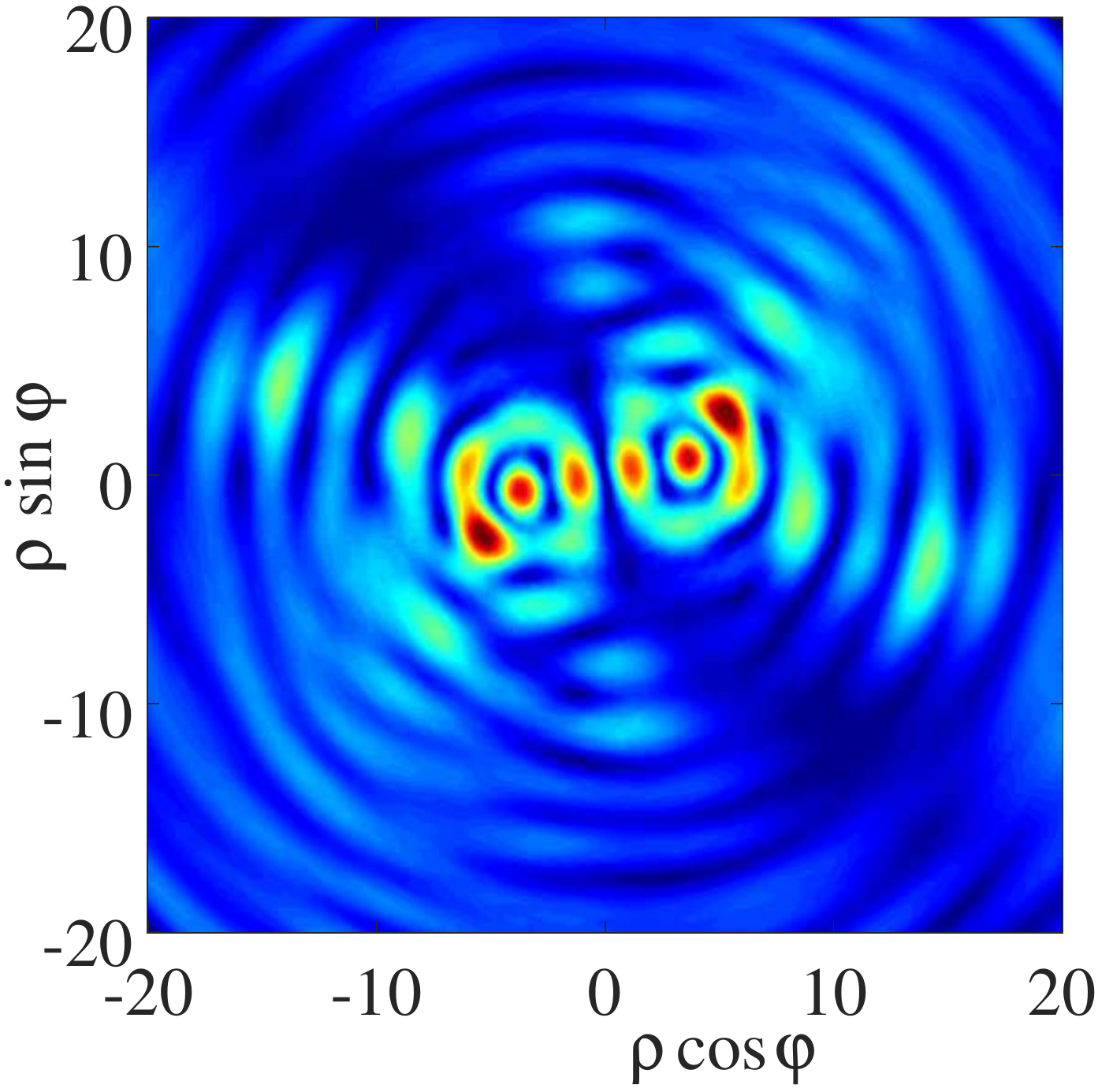}} \subfigure[]{%
\includegraphics*[width=4.0cm]{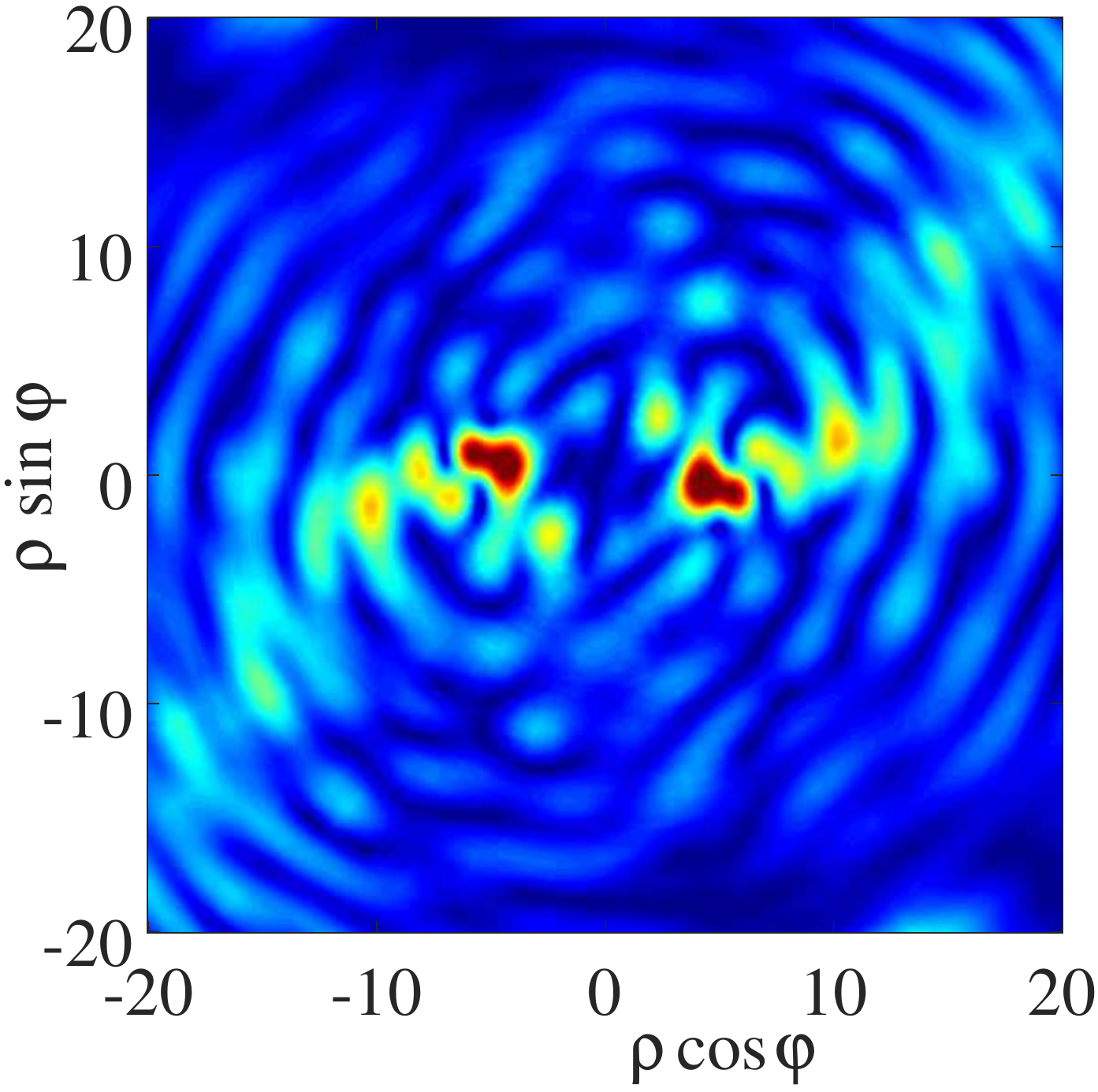}}
\end{center}
\caption{An example of the instability development, starting from the
two-lobe dissipative azimuthon, with $s=1$, and leading to establishment of
a random set of isolated hot spots. In this case, the input is similar to
that displayed in Fig. \protect\ref{Fig3}(c). Examples of a transient
regular amplitude pattern, obtained at $\protect\zeta =20.625$ (a), and of a
random pattern, obtained at $\protect\zeta =93.75$ (b). The developing
pattern is eventually fully randomized (not shown here). Parameters are $M=4$%
, $\protect\alpha =4$ for the NLSE, and $s=1$, $|b_{s}|=0.8$, $\protect\eta %
=0.04$ for the Bessel-beam input (\protect\ref{BESSEL-AZIMUTHON}).}
\label{Fig6}
\end{figure}

\subsection{Instability development scenarios}

\begin{figure}[t]
\begin{center}
\subfigure[]{\includegraphics*[width=4.0cm]{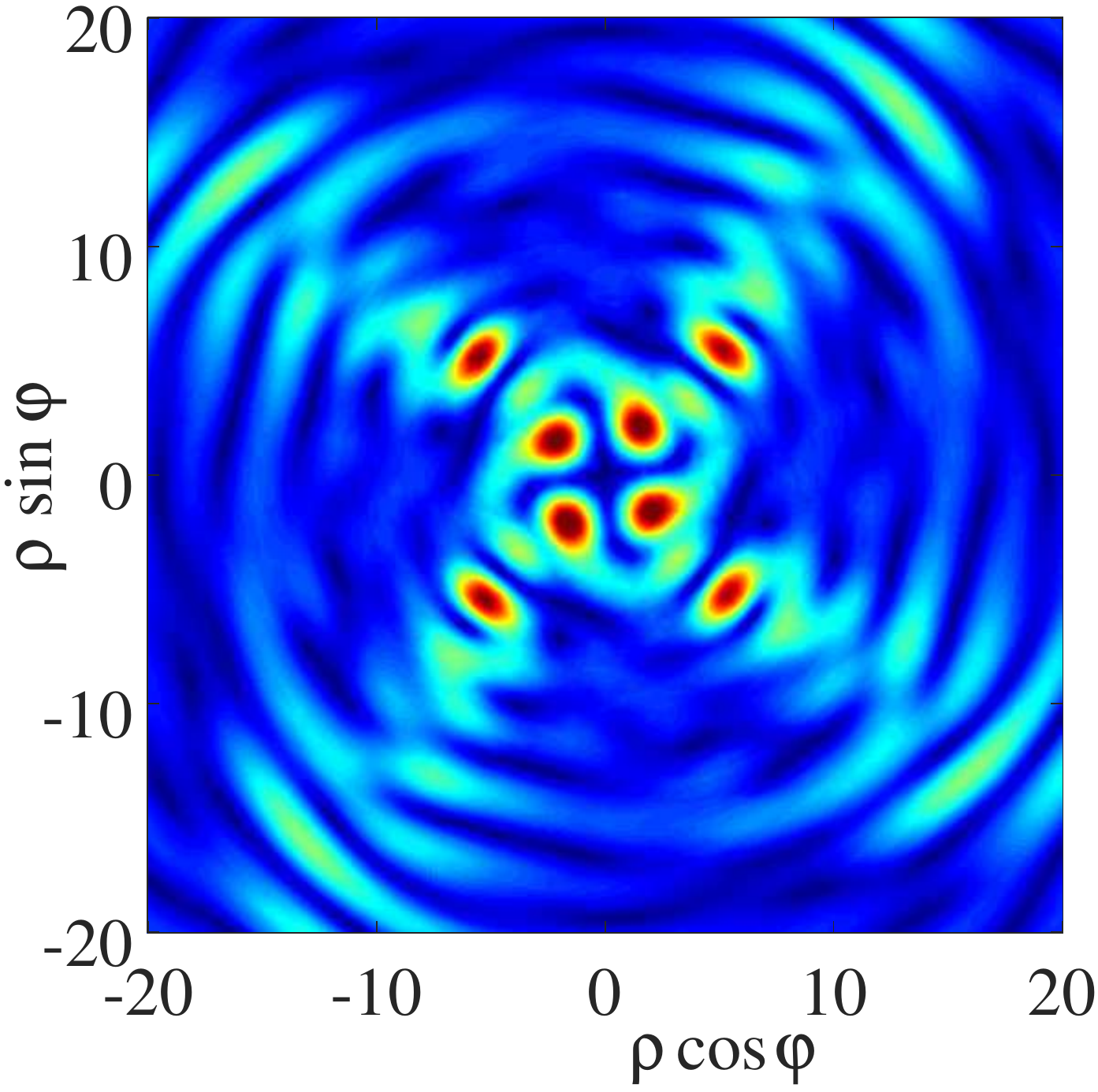}} \subfigure[]{%
\includegraphics*[width=4.0cm]{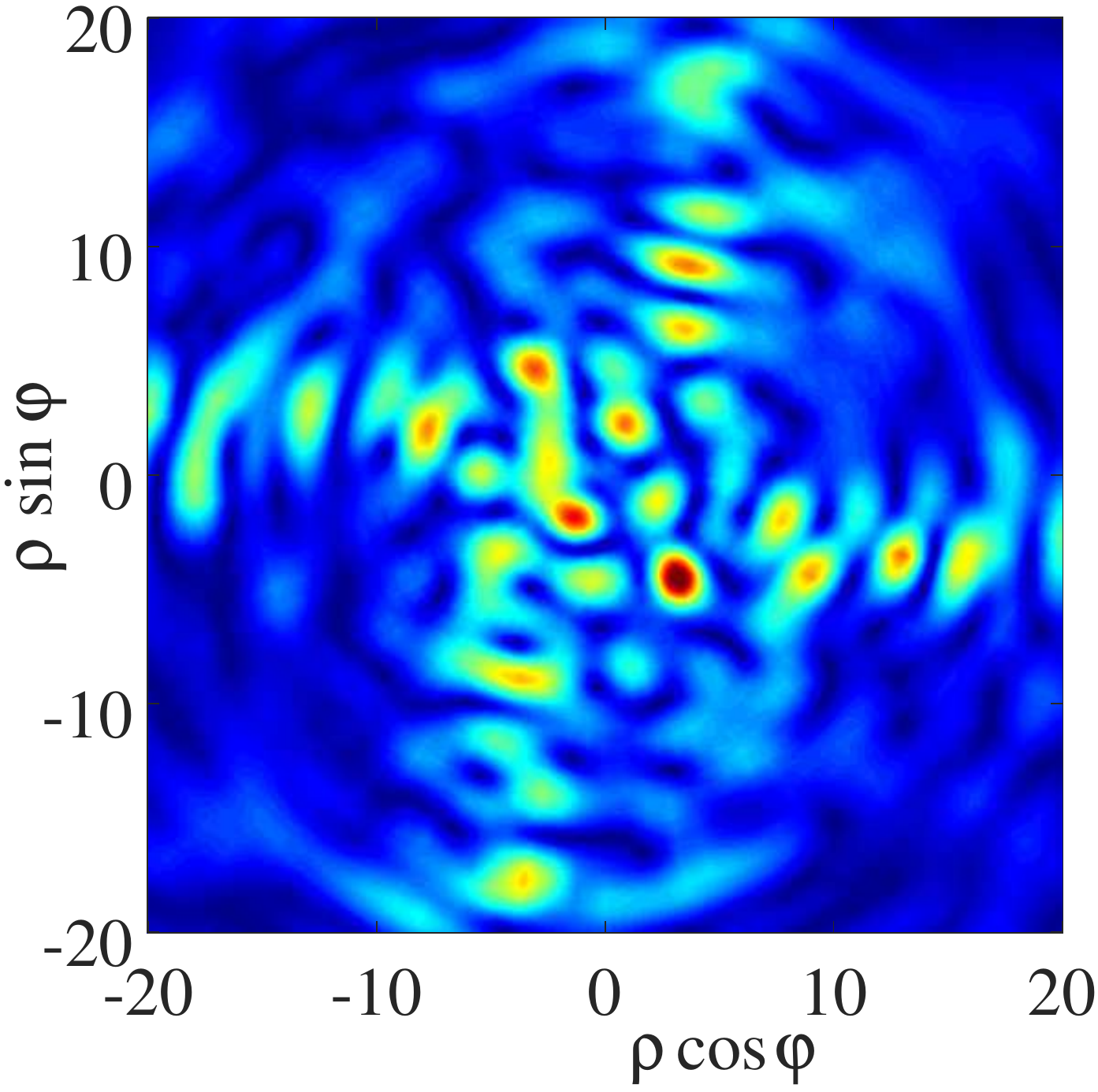}}
\end{center}
\caption{An example of the instability development, starting from the
four-lobe dissipative azimuthon, with $s=2$. In this case, the input is
similar to that displayed in Fig. \protect\ref{Fig4}(a). Examples of a
transient regular amplitude pattern, obtained at $\protect\zeta =16.875$
(a), and of a randomized one, established at $\protect\zeta =105$ (b).
Parameters are $M=4$, $\protect\alpha =4$ for the NLSE, and $s=2$, $%
|b_{s}|=0.8$, $\protect\eta =0.04$ for the Bessel-beam input (\protect\ref%
{BESSEL-AZIMUTHON}).}
\label{Fig7}
\end{figure}

\begin{figure}[b]
\begin{center}
\subfigure[]{\includegraphics*[width=4.0cm]{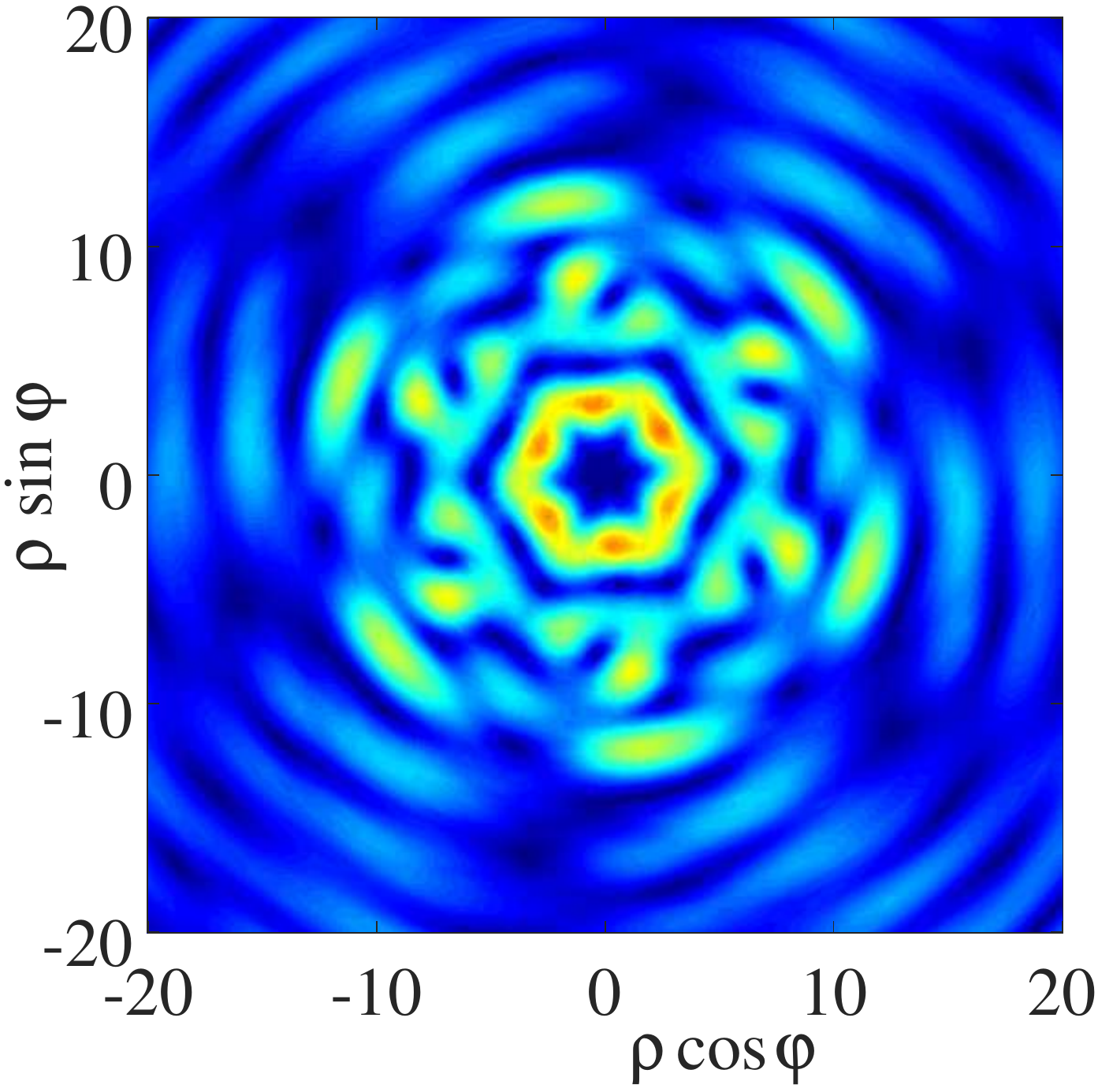}} \subfigure[]{%
\includegraphics*[width=4.0cm]{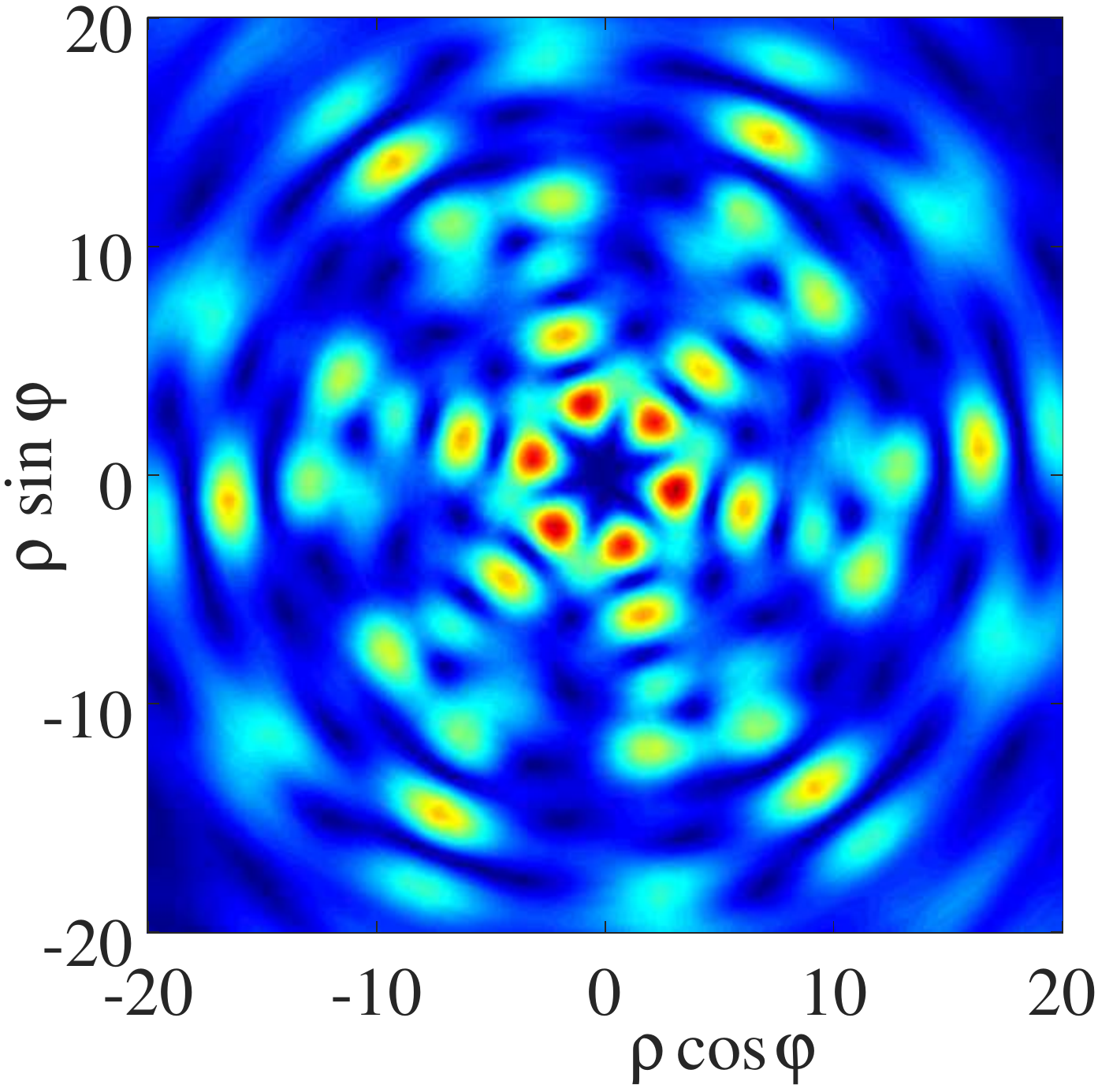}}
\end{center}
\caption{Examples of the instability development of the six-lobe dissipative
azimuthon, proceeding through a sequence of regular patterns, the input
pattern being close to that shown in Fig. \protect\ref{Fig5}(a), while the
final pattern (not shown here) is a random set of isolated hot spots.
Examples of transient regular patterns are displayed at (a) $\protect\zeta %
=7.5$ and (b) $\protect\zeta =13.125$. Parameters are $M=4$, $\protect\alpha %
=4$ for the NLSE, and $s=3$, $|b_{s}|=0.8$, $\protect\eta =0.04$ for the
Bessel-beam input.}
\label{Fig8}
\end{figure}

Direct simulations make it possible to identify two basic scenarios of the
development of unstable patterns. Close to the stability boundary, i.e., for
the values of $\alpha $ slightly exceeding $\alpha _{\mathrm{th}}$, unstable
rotating patterns spontaneously turn into oscillating ones, which keep
rotating and may remain robust by themselves. For higher values of $\alpha $%
, an unstable azimuthon decays into random patterns, by progressively
loosing its symmetry. Three examples of this dynamics are shown in Figs. \ref%
{Fig6}, \ref{Fig7} and \ref{Fig8} for $s=1,2$ and $3$, respectively. The
instability develops, at first, through a long sequence of regular patterns
featuring the same $2s$-fold symmetry as the input, which follow each other
in a random way. Then the pattern becomes less and less regular, developing
intrinsic oscillations, until getting complete random.

The instability-development scenario, demonstrated by the numerical
solutions, is essentially the same for $s=1,2$ and $3$, but the symmetry
with respect to the rotation by $\Delta \varphi =\pi $ for $s=1$ is much
more robust than the symmetry with $\Delta \varphi =\pi /2$ for $s=2$ or $%
\Delta \varphi =\pi /3$ for $s=3$. The patterns for $s=1$ are much simpler,
and, typically, the resulting random structure still keep the symmetry with
respect to the rotation by $\Delta \varphi =\pi $ [see, e.g., Fig. \ref{Fig6}%
(b)].

The phase pattern globally reflects the shape of the intensity pattern.
However, if one considers the phase as a function of azimuthal angle $%
\varphi $ along circumferences with different radii $\rho $, conspicuous
evolution in the course of the evolution in $\zeta $ can be identified.
Indeed, while the phase circulation is $2\pi s$ in the core of the input
configuration (\ref{BESSEL-AZIMUTHON}), and $-2\pi s$ just after the
topological charge has reversed its sign,
a central domain of small radius in which the phase circulation is zero
quickly emerges in the course of the evolution of the unstable azimuthon.
The radius of this domain grows until it fills the entire core of the
pattern, except for a few remaining phase dislocations, whilst the pattern
becomes random. Thus, the evolution leads to a conclusion that the core area
of the pattern does not carry vorticity anymore. Alternations of domains
with phase circulations $2\pi s$ and $-2\pi s$ still occurs at different
increasing values of the radius and, as expected, in the peripheral area (at
large radii) the alternations become identical to that of the input
Bessel-beam superposition.

\subsection{Pulsating and breathing dissipative azimuthons}

\begin{figure}[tbp]
\begin{center}
\subfigure[]{\includegraphics*[width=4.0cm]{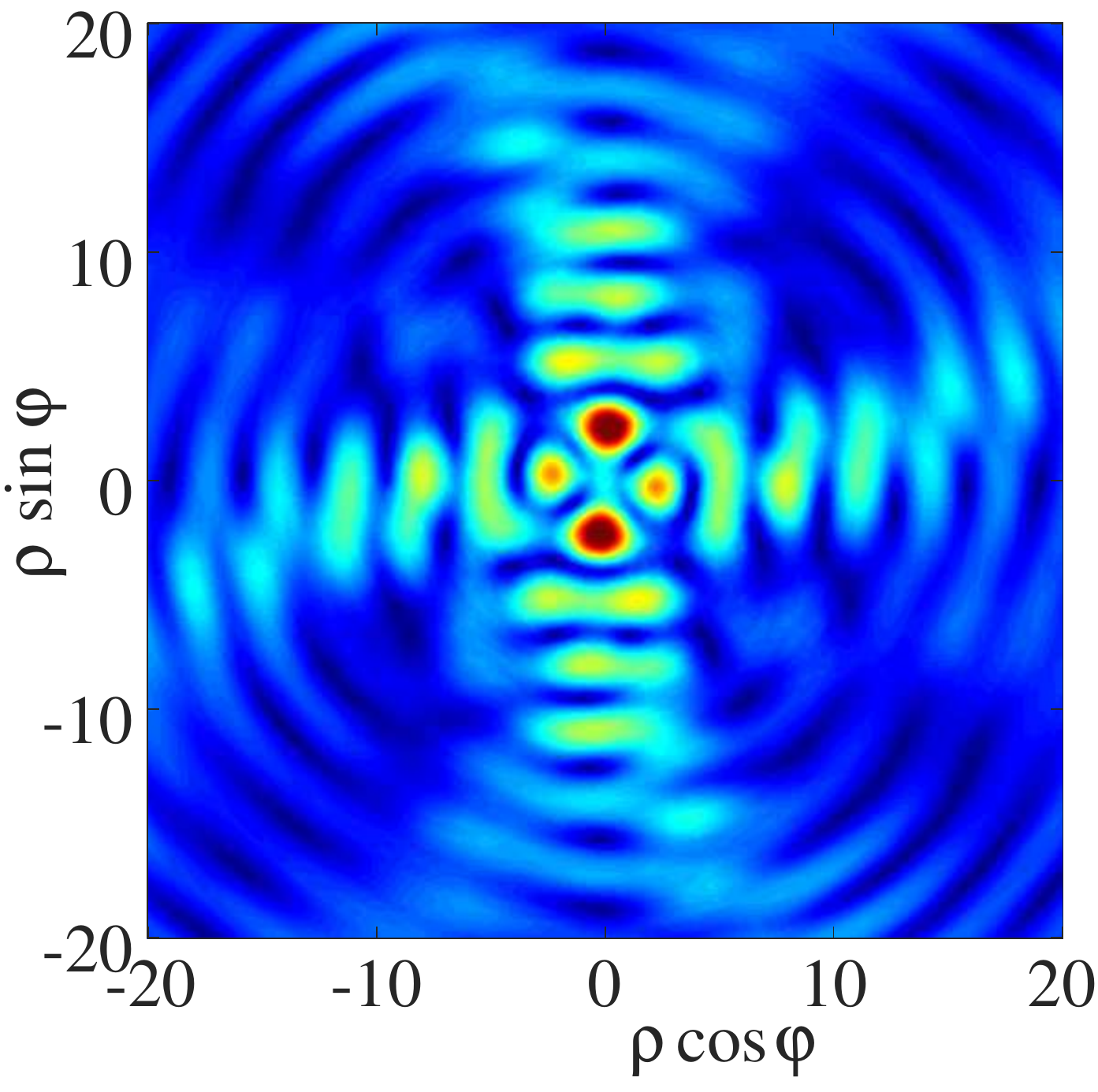}} \subfigure[]{%
\includegraphics*[width=4.0cm]{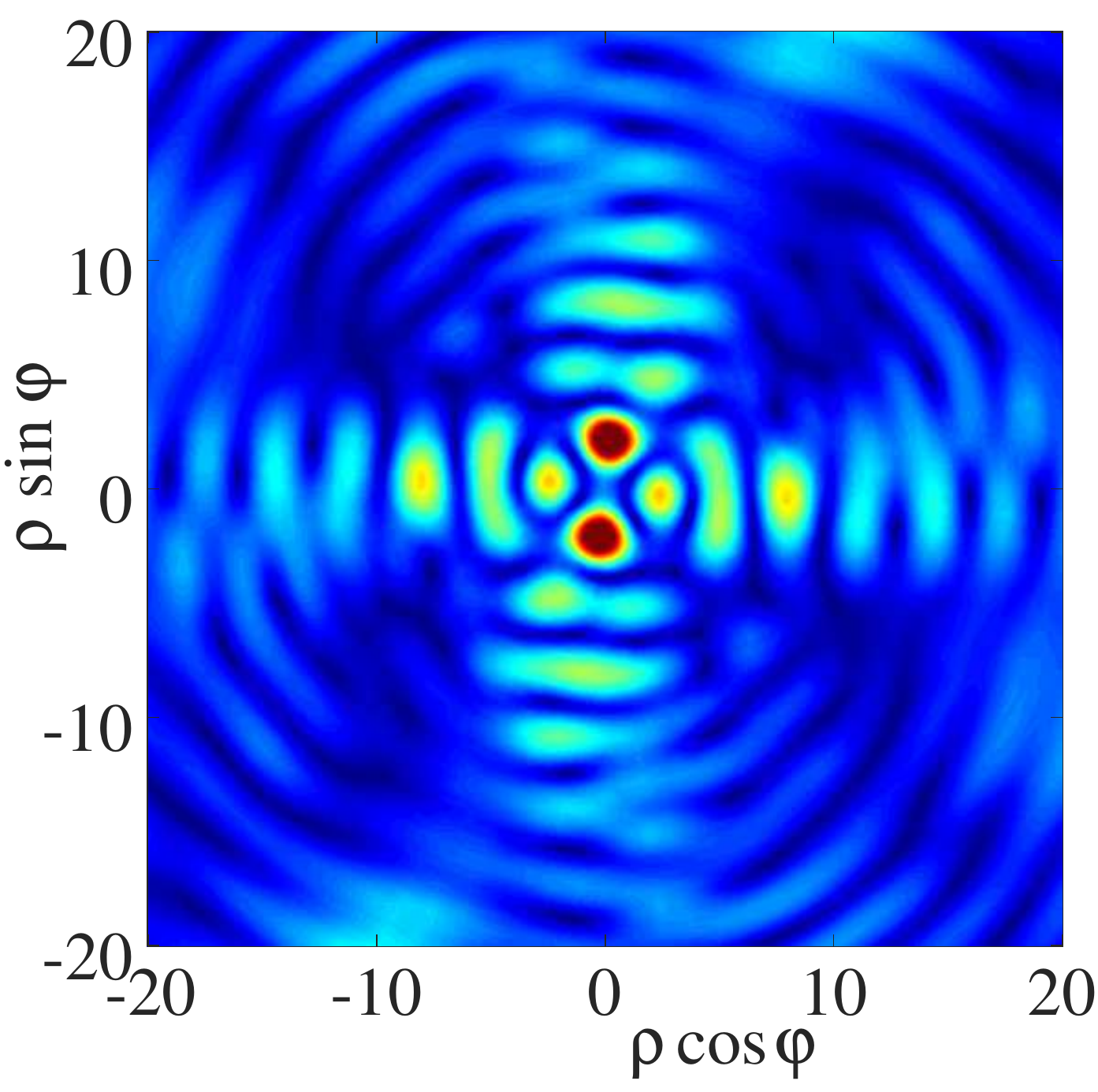}}
\end{center}
\caption{An example of the formation of a pulsating dissipative azimuthon,
triggered by the instability of a four-lobe azimuthon. In this case, input (%
\protect\ref{BESSEL-AZIMUTHON})\ is similar to that in Fig. \protect\ref%
{Fig4}. (a) At $\protect\zeta =115.625$, the instability has just led to the
appearance of a two-lobe pattern. (b)A snapshot of a persistent oscillatory
pattern at $\protect\zeta =231.25$ [it has rotated by $\Delta \protect%
\varphi =\protect\pi /2$ between (a) and (b)]. Parameters are $s=2$, $M=4$, $%
\protect\alpha =2.6$, $|b_{s}|=0.8$, and $\protect\eta =0.04$.}
\label{Fig9}
\end{figure}

In the case of vorticity $s=2$, a peculiar instability scenario exhibits
spontaneous transition of the four-lobe pattern into a two-lobe one. The
development of this scenario is quite slow, allowing the initial four-lobe
pattern to propagate considerable distances, keeping an apparently stable
shape. The transition commences with oscillations of the amplitude of two
pairs of hot spots which constitute the cross (four-lobe) structure of the
input. Such oscillations may lead directly to the destruction of the
structure, or proceed in an apparently persistent way. Due to a limited
propagation distance in the simulations and very low instability growth
rate, it is not always possible to distinguish these outcomes. An example is
produced in Fig. \ref{Fig9} for $s=2$, $|b_{s}|=0.8$, and $\alpha =2.6$, in
which the emerging oscillatory state appears to be stable. In this case,
input (\ref{BESSEL-AZIMUTHON}) quickly forms a dissipative azimuthon, that
propagates without any visible instability up to $\zeta \simeq 75$. Then,
amplitudes of the lobes start to oscillate until the pattern reshapes into a
fully robust two-lobe structure.

\begin{figure}[tbp]
\begin{center}
\subfigure[]{\includegraphics*[width=4.0cm]{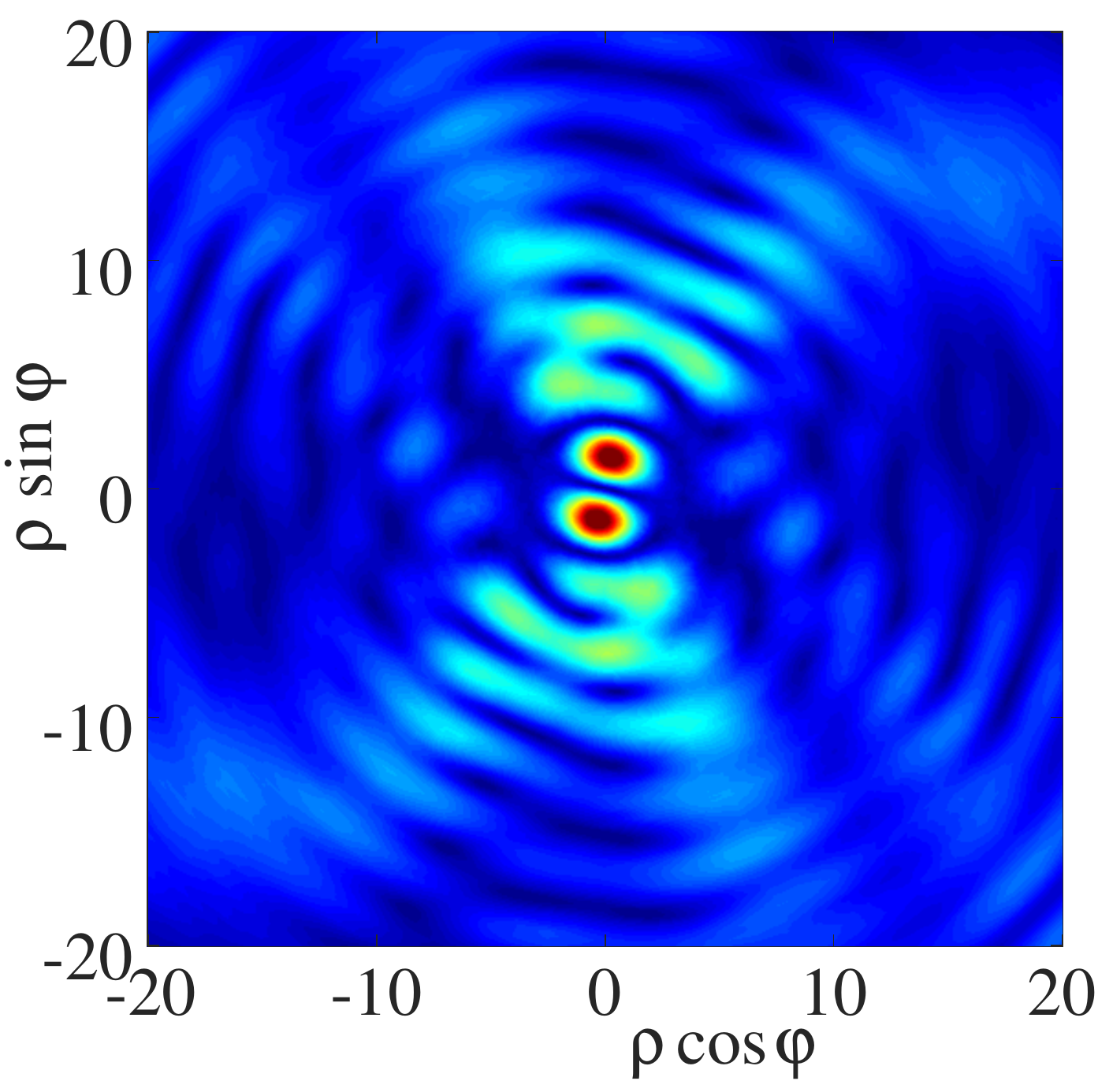}} \subfigure[]{%
\includegraphics*[width=4.0cm]{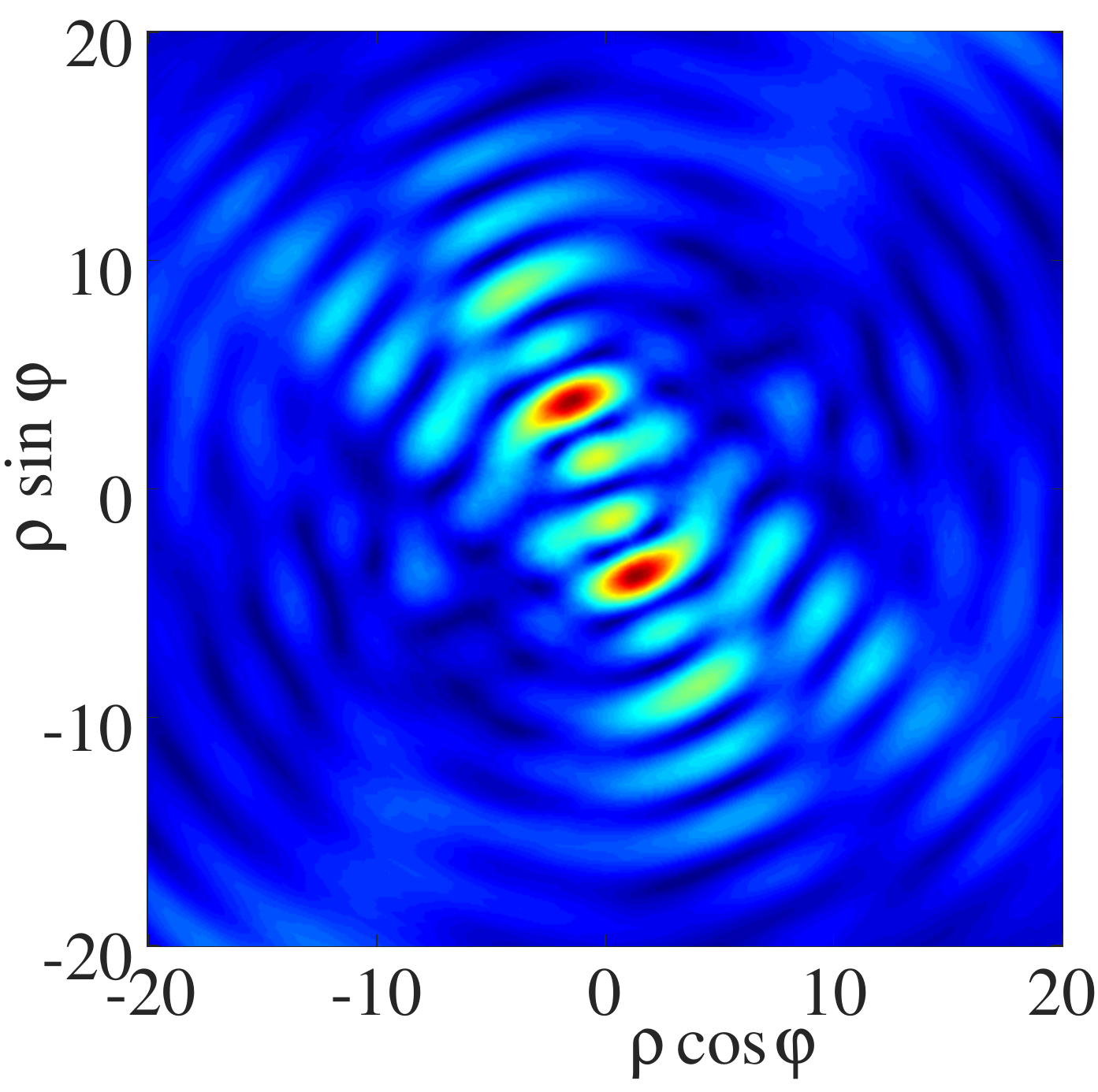}}
\end{center}
\caption{A breathing dissipative azimuthon. It oscillates between two
configurations: at $\protect\zeta =193.75$ (a) and $200$ (b), the major
amplitude maxima are located, respectively, at internal and external
positions. Parameters are $s=1$, $M=4$, $\protect\alpha =3$, $|b_{s}|=0.6$,
and $\protect\eta =0.04$. See Supplemental Material [URL of Fig10.avi] for a
detailed picture of the breathing-azimuthon dynamics.}
\label{Fig10}
\end{figure}

Another noteworthy propagation regime triggered by the instability for $s=1$
is the formation of a permanently breathing dissipative azimuthon. As seen
in Fig. \ref{Fig10}, in this case oscillations occur between a pattern in
which the hot spots with the largest amplitude are closest to the center
[Fig. \ref{Fig10}(a)], and another one, in which the hot spots are located
on a secondary ring [Fig. \ref{Fig10}(b)]. This regime sets in as a
persistent one, after a transient stage in which the intensity maxima of the
farther separated spots are lower. The dynamics of the breathing dissipative
azimuthon is displayed in Supplemental Material [URL of Fig10.avi].

\begin{figure}[t]
\begin{center}
\subfigure[]{\includegraphics*[width=4.0cm]{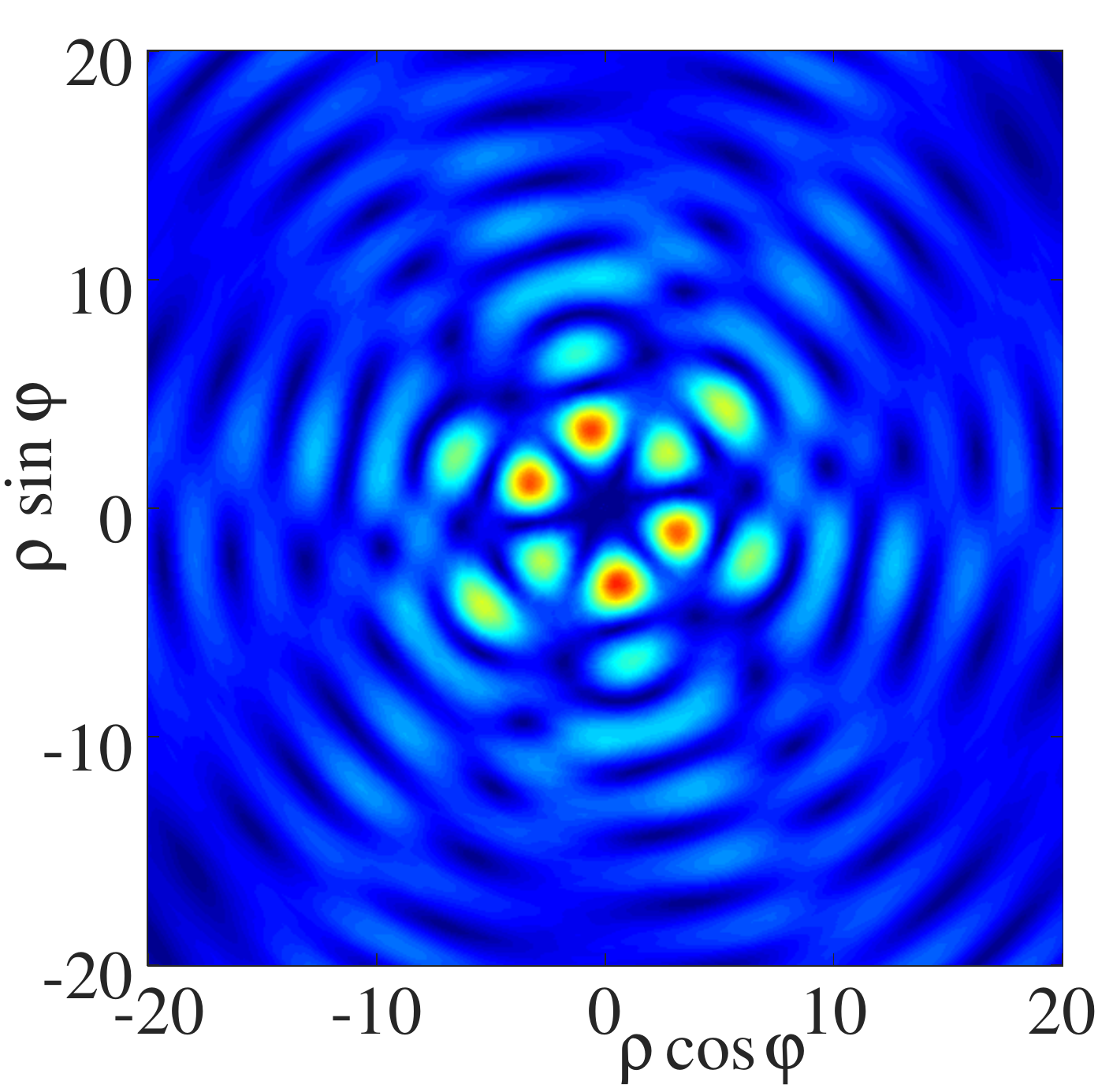}} \subfigure[]{%
\includegraphics*[width=4.0cm]{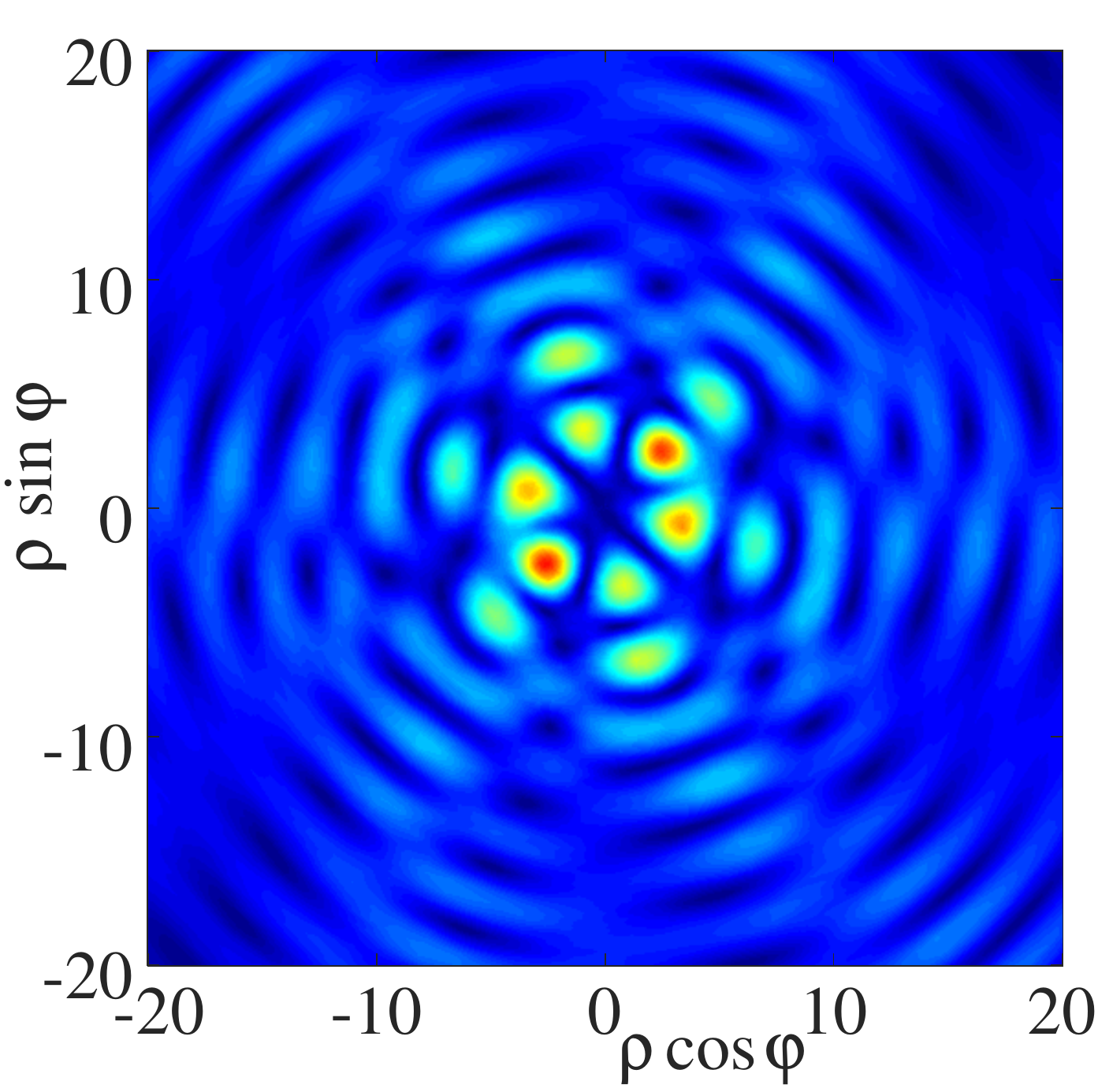}} \subfigure[]{%
\includegraphics*[width=4.0cm]{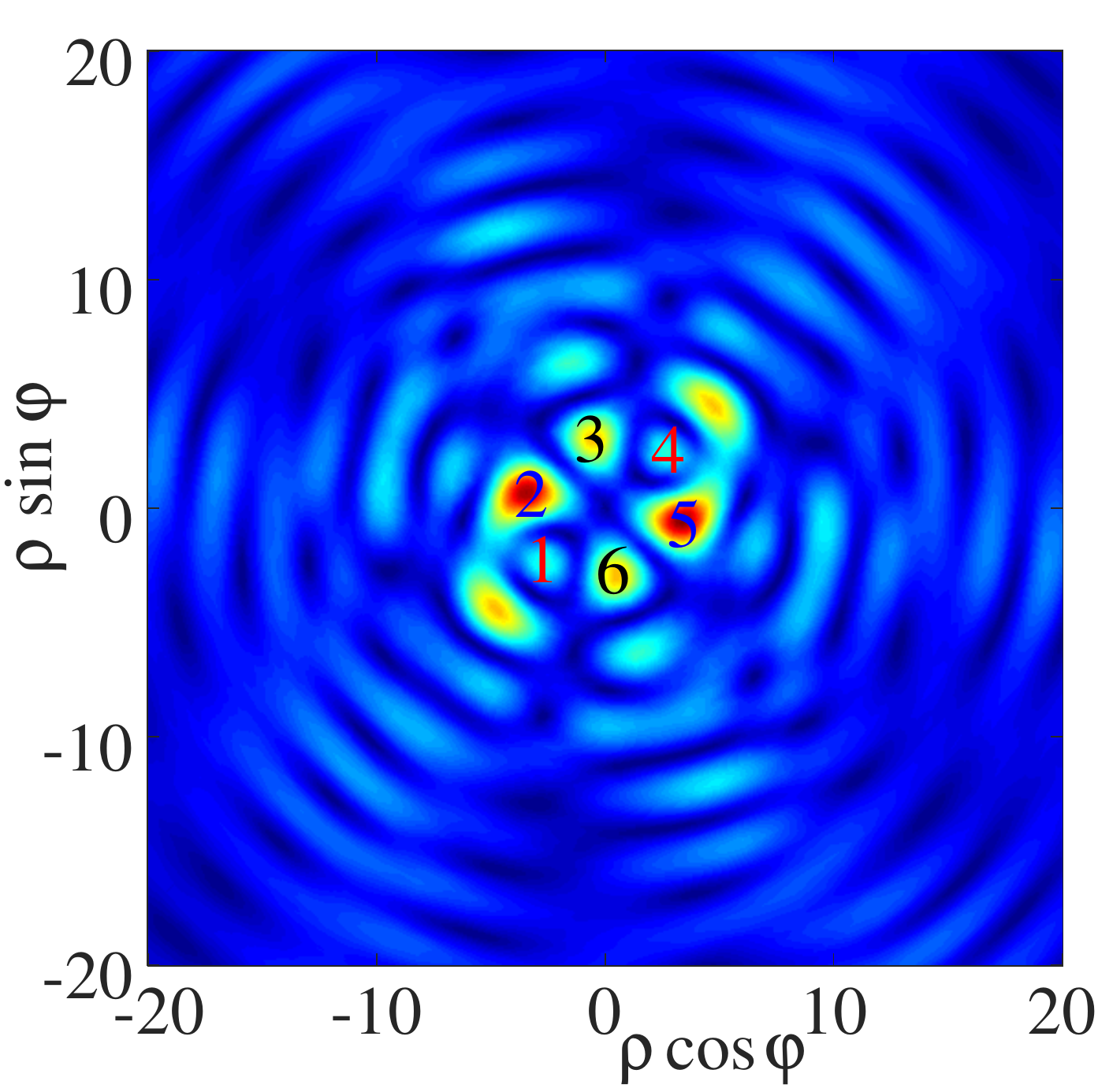}} \subfigure[]{%
\includegraphics*[width=4.0cm]{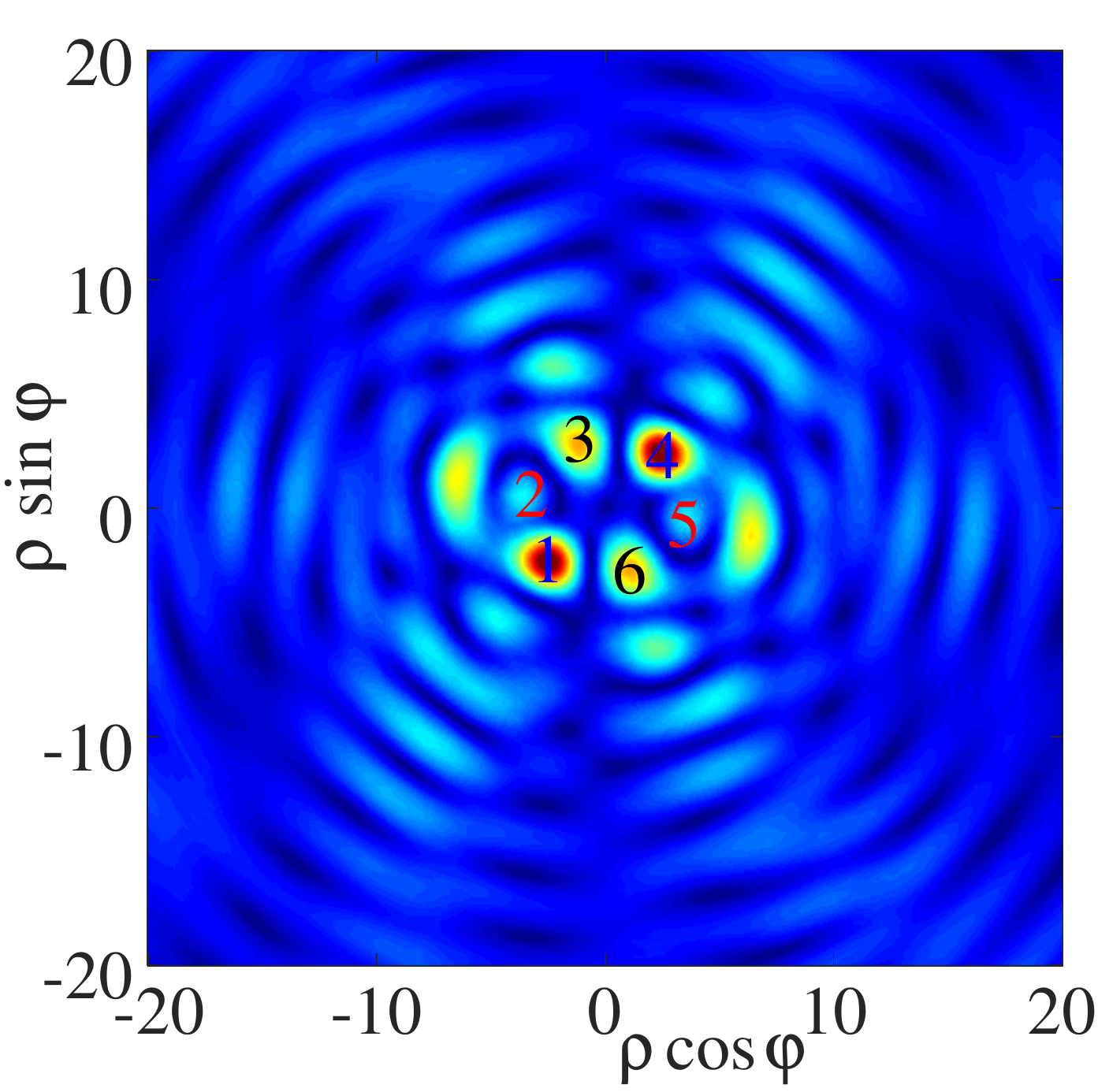}}
\end{center}
\caption{An example of the unstable evolution of the six-lobe dissipative
azimuthon. The respective input (\protect\ref{BESSEL-AZIMUTHON}) is similar
to that presented in Fig. \protect\ref{Fig5}. At the first stage of the
evolution, the pattern oscillates between configurations shown in (a), at $%
\protect\zeta =125$, and (b), at $\protect\zeta =128.125$. At the second
stage, the pattern oscillates between the configurations shown in (c), at $%
\protect\zeta =165.625$, and (d), at $\protect\zeta =168.750$. Parameters
are $s=3$, $M=4$, $\protect\alpha =3$, $|b_{s}|=0.6$, and $\protect\eta =0.04
$.}
\label{Fig11}
\end{figure}

There is also a dynamical regime exhibiting spontaneous transformation of a
transient oscillatory state, which keeps a high level of symmetry, into
another one with a lower symmetry, as shown in Fig. \ref{Fig11}. The initial
six-lobe azimuthon is similar to the one displayed Fig. \ref{Fig5} but for a
larger scaled strength of the Kerr nonlinearity, $\alpha $ [see Eq. (\ref%
{alpha})]. At the first, relatively long, stage of the evolution ($\zeta
\lesssim 75$), the pattern remains unchanged in the rotating reference
frame, before the instability commences. Then, the first oscillatory state
emerges, see Figs. \ref{Fig11}(a) and (b), which keeps essential symmetry:
the initial invariance with respect to the rotation by $\Delta \varphi =\pi
/3$ is lost, being reduced to the invariance with $\Delta \varphi =\pi $,
but the axial symmetry is conserved. The amplitude oscillates between
configurations with two opposite spots and four spots. Then, the structure
switches into a second oscillatory state, which keeps solely the symmetry
with respect to the rotation by $\Delta \varphi =\pi $, see Figs. \ref{Fig11}%
(c) and (d). In this state, oscillations occur between opposite spots in two
pairs, labeled $\left( 1,4\right) $ and $\left( 2,5\right) $ in the figure,
while the spots belonging to the third pair, $\left( 3,6\right) $, keep a
low intensity. Note that the apparent rotation by $\pi /6$, which relates
Figs. \ref{Fig11}(c) and (d), is actually a consequence of the oscillations.
Indeed, the rotation angle per se between the configurations in panels (c)
and (d), separated by propagation distance $\Delta \zeta =3.125$, is $\Delta
\varphi =\varpi \Delta \zeta =(2\eta /s)\Delta \zeta $ [see Eq. (\ref{omega}%
)], which yields $\Delta \varphi =4.77^{\circ }$.

\section{Conclusion}

We have reported the existence of a novel type of azimuthons, which
represent the propagating optical field with a stationary intensity pattern
in a uniformly rotating reference frame in the Kerr medium with nonlinear
loss, induced by multiphoton absorption in the material. Similar to
nonlinear Bessel fundamental \cite{PORRASPRL,POLESANA} and vortical \cite%
{PORRASJOSAB,PORRASPRA1} beams and ``dissipatons" \cite%
{PORRASPRA2}, the steady propagation of the rotating azimuthons in the lossy
medium is maintained by the flux from the peripheral reservoir, which stores
an indefinitely large amount of power in the slowly decaying tails of the
beam. Unlike conservative azimuthons \cite{DESYATNIKOV2,MINO}, the number $N$
of ``hot spots" (intensity maxima) and the vorticity of the
input are linked by $N=2s$, rather than being mutually independent.

The rotating dissipative azimuthons are excited by the coherent
superposition of two Bessel beams with opposite topological charges and
slightly different cone angles, cf. Refs. \cite{VASILYEU,ROP}. In comparison
to the non-rotating intensity patterns excited by the Bessel-beam pairs with
identical conicities, the rotating azimuthons form faster and are more
robust. The existence and stability of these modes in the self-focusing Kerr
medium is provided by the nonlinear absorption. If the absorption is turned
off, the input superposition of the Bessel beams does not result in
formation of any stationary pattern. Unstable rotating azimuthons in this
model are interesting objects too, because the development of the
instability gives rise to various dynamical regimes, including persistently
pulsating and breathing azimuthons, as well as the transition to
``turbulent" patterns.

These results may help to understand physics underlying the recently
observed helical filaments, excited by superpositions of Bessel-Gauss beams
with opposite vorticities in air and CS$_{2}$ \cite{BARBIERI,LU}, where the
interplay of the Kerr nonlinearity and multiphoton absorption, induced by
ionization of air, plays a key role in the propagation.

\section*{Acknowledgements}

M.A.P. acknowledges funding by Spanish Ministerio de Econom\'{\i}a y
Competitividad, grant No. PGC2018-093854-B-I00 and No. FIS2017-87360-P.
B.A.M. appreciates support from the Israel Science Foundation, through grant
No. 1286/17.

\appendix*

\section{Numerical schemes}

The validity of all the  numerical results reported above has been verified by running the simulations independently
in the Cartesian and polar coordinates, by means of different algorithms, with different discretization meshes. In the former case, a standard symmetrized split-step Fourier method applied to the rectangular mesh, using the fast Fourier transform to perform linear steps of the integration, and a trapezoidal method for the nonlinear steps. With the polar coordinates, a $5$-points finite-difference scheme was employed to evaluate the transverse derivatives, and a $4$th-order Runge-Kutta scheme for the axial integration on the polar mesh. It has been
concluded that all the results produced by means of both coordinates systems are identical.

Special boundary conditions were required to simulate the propagation of Bessel-beam superpositions with tails slowly
decaying at $r\to\infty$. In all the simulations, the evolving field was set equal to
the linearly propagated field, whose analytical expression is given by Eq. (\ref{lin-Bessel}), in a narrow strip attached to the transverse computational boundary, which contains a few radial points of the discretization mesh. From the physical point of view, these boundary conditions are justified as the propagation of the small-amplitude tails is initially linear, and the boundary conditions emulate the divergence of the total power in the reservoir. On the computational side, the same boundary conditions initially eliminate unphysical inward-propagating boundary waves, that would emerge if zero boundary conditions were used. Results of the simulations become invalid at values of the propagation distance at which nonlinear excitations propagating from the high-intensity beam's core outwards are reflected at the strip boundary and come back to the core.

\end{document}